\documentclass[fleqn,usenatbib]{mnras}

\usepackage{newtxtext,newtxmath}

\usepackage[T1]{fontenc}
\usepackage{ae,aecompl}

\usepackage{graphicx}	
\usepackage{amsmath}	
\usepackage{subfig}
\usepackage{amsmath, epsfig,natbib}
\usepackage{multicol}
\usepackage{color,ulem}
\usepackage{newtxtext,newtxmath}
\definecolor{webgreen}{rgb}{0,.5,0}
\definecolor{webbrown}{rgb}{.6,0,0}

 \newcommand{\kms}{\mbox{$\>{\rm km\, s^{-1}}$}}
\newcommand{\pc}{\>{\rm pc}}
\newcommand{\kpc}{\mbox{$\>{\rm kpc}$}} 

\newcommand{\Gyr}{\mbox{$\>{\rm Gyr}$}}
\newcommand{\Myr}{\mbox{$\>{\rm Myr}$}}

\newcommand{\Msun}{\>{\rm M_{\odot}}}

\newcommand\degrees{^\circ}

\title [Bar destruction in minor mergers ] 
{Fate of stellar bars in minor merger of galaxies }
\author[Ghosh et al.]
	{Soumavo Ghosh$^{1}$\thanks{E-mail : soumavo@iucaa.in},
	      Kanak Saha$^{1}$,
	Paola Di Matteo$^{2}$,
	Francoise Combes$^{3}$\\
$^1$  Inter-University Centre for Astronomy and Astrophysics, Pune 411007, India\\
$^2$  GEPI, Observatoire de Paris, PSL Research University, CNRS, Sorbonne Paris Cité, 5 place Jules Janssen, 92190, Meudon, France\\
$^3$  Observatoire de Paris, LERMA, College de France, CNRS, PSL University, Sorbonne University, Paris, France\\
} 
\date{Accepted 2021 January 25. Received 2021 January 25; in original form 2020 August 10}

\pubyear{2021}

\begin{document}
\label{firstpage}
\pagerange{\pageref{firstpage}--\pageref{lastpage}}
\maketitle


\begin{abstract} 
Minor merger of galaxies  are common during the evolutionary phase of galaxies. 
Here, we investigate the dynamical impact of a minor merger (mass ratio 1:10) event on the {\it final} fate of a stellar bar in the merger remnant. To achieve that, we choose a set of minor merger models from the publicly available GalMer library of galaxy merger simulations. The models differ in terms of their orbital energy, orientation of the orbital spin vector, and morphology of the satellite galaxy (discy/spheroidal). We demonstrate that the central stellar bar, initially present in the host galaxy, undergoes a transient bar amplification phase after each pericentre passage of the satellite; in concordance with past studies of bar excitation due to tidal encounter.  However, once the merger happens, the central stellar bar weakens substantially in the post-merger remnants. The accumulation of satellite's stars in the central region of merger remnant plays a key role in the bar weakening process; causing a net increase in the central mass concentration as well as in the specific angular momentum content. 
We find that the efficiency of mass accumulation from the satellite in the central parts of merger remnants depends on the orbital parameters as well as on the satellite's morphology. Consequently, different minor merger models display different degrees of bar weakening event. This demonstrates that minor merger of galaxies is a plausible avenue for bar weakening in disc galaxies.
\end{abstract}




\section{Introduction}
\label{sec:intro}
In the Lambda cold dark matter (LCDM) paradigm, galaxies grow hierarchically via major mergers and/or multiple minor mergers, and accretion of cold gas \citep{WhiteandRees1978,Fall1980}. The minor merger of galaxies (mass ratio greater than or equals to 1:10) are shown to be less catastrophic than the major merger (mass ratios from 1:1 to 1:3) events, so that they can preserve the disc morphology in the post-merger remnant. However, the details of maintaining a kinematically-cold thin disc and a kinematically-hotter thick disc in the merger remnant depends on the fraction of interstellar gas present in the merging disc galaxies \citep[e.g., see][]{Villalobosetal2008,Mosteretal2010}. Minor mergers can happen frequently in the local Universe \citep[e.g., see][]{Frenketal1988,CarlbergandCouchman1989,LaceyandCole1993,Gaoetal2004,Jogeeetal2009,Kavirajetal2009}. Therefore, it is of key interest to understand the detailed role of minor merger of galaxies in driving the evolution of disc galaxies.

In the past, both theoretical and observational efforts have focused on the impact of minor mergers on galaxy's evolution and reshaping their kinematics. Minor mergers are shown to leave a number of characteristic morphological finger-prints in disc galaxies \citep[e.g., see][]{Ibataetal1994,Ibataetal2001,Yannyetal2003,Erwinetal2005,Ibataetal2005,Youngeretal2007,Feldmanetal2008,Kazantzidisetal2009}. This also causes heating of the disc and thickening the disc in the vertical direction \citep{Quinnetal1993,Walkeretal1996,VelazquezandWhite1999,Fontetal2001,Kazantzidisetal2008,Quetal2011a}, decreasing the specific angular momentum of stellar disc in the post-merger remnant, irrespective of the orbital configuration or the morphology of satellite galaxy \citep{Quetal2010,Quetal2011b}, producing inner components (such as inner ring, inner disc etc.) for unbarred galaxies \citep{Eliche-Moraletal2011}, enhancing star formation activities \citep[e.g., see][]{Kaviraj2014}, radially distributing the chemical abundances in Milky Way-like galaxies \citep[e.g., see][]{Zinchenkoetal2015}, and transferring angular momentum to the dark matter halo via action of stellar bars \citep{Debattistaetal2006,SellwoodandDebattista2006}.

Past observations have shown that about two-third of the disc galaxies in the local Universe host bars \citep[e.g., see][]{Eskridgeetal2000,Whyteetal2002,Aguerrietal2009,Mastersetal2011}. The occurrence of bars is found to depend strongly on the stellar mass \citep[e.g.,][]{NairandAbraham2010}, Hubble type \citep[e.g.,][]{Aguerrietal2009,Butaetal2010,NairandAbraham2010} of the host galaxies. Whether the remaining one-third of disc galaxies in the local Universe are hostile to the bar formation and their growth, or the bar has been destroyed during their evolutionary trajectory -- it is still not completely understood \citep{SahaElmegreen2018}. Destroying completely a central stellar bar has proven to be an arduous task. Past theoretical studies have identified central mass concentration and inflow of gas as plausible mechanisms for bar destruction; however it might require prodigious amount of gas inflow or a very high central mass concentration \citep[e.g., see][]{Pfenniger1990,ShenSellwood2004,Athanassoulaetal2005,Bournaudetal2005,HozumiHernqusit2005,Athanassoulaetal2013}. Also, recent observational work by \citet{Pahwa2018} showed the presence of prominent bars in several low-surface-brightness (LSB) galaxies with high gas fraction. The bar fraction in the high-redshift galaxies is still debated; some studies claimed a decreasing bar fraction with increasing redshift \citep[e.g., see][]{Shethetal2008,Melvinetal2014,Simmonsetal2014}, while some other studies showed a constant bar fraction up to redshift $z \sim$ 1 \citep[e.g.,][]{Elmetal2004,Jogeeetal2004}. Nevertheless, this large abundance of stellar bars in disc galaxies and the relatively larger frequency of occurrence of minor merger events \citep[e.g.,][]{FakhouriandMa2008} of disc galaxies raises an important question -- what happens to a stellar bar when the host galaxy experiences a minor merger event with a satellite galaxy?

Past studies have focused on the dynamical effect of tidal encounter in triggering the bar instability in disc galaxies. The increased bar fraction in the central regions of Virgo and and Coma cluster suggested that tidal interactions can trigger bar formation in disc galaxies, especially in the Early-type disc galaxies \citep[e.g., see][]{Thompson1981,Giuricinetal1993,Anderson1996,Barazzaetal2009,M_ndez_Abreu_2012,Linetal2014}. Later observational studies indicated that the bar formation due to a fly-by encounter depends on the galaxy's mass and their ability to maintain a cold disc component against the heating caused the tidal encounter \citep[e.g.,][]{M_ndez_Abreu_2012}. Bar formation due to a tidal encounter and its effect on the bar properties has been further studied extensively using $N$-body simulation of disc galaxies  \citep[e.g., see][]{Noguchi1987,Gerinetal1990,Sundinetal1993,MiwaandNoguchi1998,Aguerri2009,Peiranietal2009,Langetal2014,Inmaetal2017}. Additionally, the properties of the resulting bar are shown to depend on the presence of the interstellar gas \citep[e.g.,][]{Berentzenetal2003}, mass ratio of the galaxies, and/or the relative phase of the bar and the companion at pericentre \citep[e.g., see][]{Gerinetal1990,Sundinetal1993,Langetal2014,Lokasetal2014}.
Despite a significant research in the field, the exact dynamical role of {\it minor mergers} on the {\it final} fate of a stellar bar remains to be explored . This is particularly true when the companion/satellite galaxy ultimately plunges into the host galaxy and the host galaxy readjusts after the merger is completed. The exact role of different orbital parameters, Hubble type of the companion, gas fraction in disc galaxy is not known either in context of reshaping the $m=2$ bar mode during a minor merger event.

In this paper, we carry out a systematic study of the temporal evolution of bar properties and the associated disc kinematics during a minor merger event while varying different orbital parameters, nature of satellite galaxies. For this, we make use of the publicly available minor merger simulation models from the GalMer database \citep{Chillingarianetal2010}. This library offers to study the physical effects of minor merger process, encompassing a wide range of cosmologically motivated initial conditions; thus, it is well-suited for the goal of this paper. 

The rest of the paper is organised as follows:\\
  Section~\ref{sec:simu_setup} provides the details of minor merger models used here while Section~\ref{sec:temp_bar} quantifies the temporal change of the stellar bar in minor mergers. Section~\ref{sec:physics_barweakening} provides the details of underlying physical mechanisms liable for bar weakening. Section~\ref{sec:satellite_dependence} discusses the dependence on the morphology of the satellite galaxy. Sections~\ref{sec:discussion} and \ref{sec:conclusion} contain discussion and  the main findings of this work, respectively.

\section{Models of minor mergers -- GalMer database}
\label{sec:simu_setup}

GalMer \footnote{ available on \href{http:/ /galmer.obspm.fr} {http://galmer.obspm.fr}} is a publicly available library of $N$-body+smooth particle hydrodynamics (SPH) simulation of  galaxy mergers to probe the details of galaxy formation through hierarchical merger process. The morphology of galaxy models ranges from ellipticals to late-type, gas-rich spirals. A galaxy model consists of a non-rotating spherical dark matter halo, a stellar and a gaseous disc (optional), and a central non-rotating bulge (optional). The dark matter halo and the central bulge (if present) are modelled as Plummer sphere \citep{Plummer1911} whereas the stellar and the gaseous disc are modelled as Miyamoto-Nagai density profiles \citep{Miyamoto-Nagai1975}.

GalMer offers three different galaxy interaction/merger scenarios, namely, the {\it giant-giant} major merger (mass ratio of 1:1), {\it giant-intermediate} merger (mass ratio of 1:2), and the {\it giant-dwarf} minor merger (mass ratio of 1:10). The total number of particles ($N_{tot}$) varies from giant-giant interaction ($N_{tot}$ = 120, 000) to giant-dwarf interaction ($N_{tot}$ = 480, 000). The orientation of each galaxy in the orbital plane is characterised by the spherical coordinates, $i_1$, $i_2$, $\Phi_1$, and $\Phi_2$ \citep[for details, see fig.~3 of ][]{Chillingarianetal2010}. The GalMer suite provides only one orbital configuration for the  giant-dwarf interaction, characterised by $i_1 =33^{\circ}$ and $i_2 =130^{\circ}$. We note that this choice is in compliance with the expectation for a random distribution of inclinations between halo spins and orbital planes. Past study by \citet{KhochfarandBurkert2006}, using a high-resolution cosmological simulation, showed that the distribution of the angle between the spin plane of the halo and the orbital plane of the satellite follows a sinus function; thus, justifying our choice of $33^{\circ}$ inclination \citep[for further details, see][]{Chillingarianetal2010}. The impact parameter ($b$) and hence the initial angular momentum of our orbital set up is another key parameter deciding the outcome of the minor merger interaction. Since the probability of interaction is proportional to $\pi b^2$, a very large impact parameter would simply delay the merger and too small an impact parameter (e.g., radial orbit in the limit $b \rightarrow 0$) is less probable. So we estimated as the most probable radius range the virial radius ($r_{200}$) following standard cosmological parameters (with $H_{0}=70$ km s$^{-1}$ Mpc$^{-1}$) for the host galaxy model gSa having a total mass of $M=2.3 \times 10^{11} M_{\odot}$ (see Table~\ref{table:key_param}). The virial radius for this host galaxy is $r_{200} = 127$~kpc. On the other hand, considering the range of initial velocities of the perturber (see Table~\ref{table:key_param}), the total energy $E_{ini} \ge 0$ (i.e., condition for being unbound) corresponds to a range of radii around $\sim 93 - 102$~kpc. In fact, the total energy could be just above zero since dynamical friction would eventually help capturing the companion. Therefore, the range of maximum probability to have an impact is indeed around $93 - 127$~kpc for the galaxy model considered here and our choice of $b=100$~kpc ($\simeq 33 R_d$) is in the right range to be representative for this model \citep[see also][]{Villalobosetal2008}.    

Following the prescription of \citet{MihosandHernquist1994}, the gas particles in the simulation are treated as `hybrid particles'. In this scheme, these hybrid particles are characterised by two masses, namely, the gravitational mass, $M_i$ which remained fixed during the simulation, and the gas mass, $M_{i, gas}$ (changing with time) which denotes the gas content of the particles. Gravitational forces are always calculated using the gravitational mass, $M_i$ while the hydrodynamical quantities make use of the time-varying gas mass, $M_{i, gas}$. If the gas fraction of a certain `hybrid particle' drops below 5 per cent of its initial gas content, then the `hybrid particle' is converted into a star-like particle while the remaining (small) amount of gas still present is distributed in the neighbouring particles \citep[for details see][]{DiMatteoetal2007}. A suitable empirical relation to follow the star formation process is implemented so as to reproduce the observed Schmidt law for the interacting galaxies. The simulation models also include the recipes for the (gas phase) metallicity evolution as well the supernova feedback. The gas mass for the `gSa' model is $9.2 \times 10^9 \Msun$ (10 per cent of the total stellar mass) whereas the gas mass for the `gSb' model is $4.6 \times 10^9 \Msun$ \citep[20 per cent of the total stellar mass; for details see][]{Chillingarianetal2010}.

The merger simulations are evolved using a TreeSPH code by \citet{SemelinandCombes2002}. The gravitational forces are calculated using a hierarchical tree method \citep{BarnesandHut1986} with a tolerance parameter $\theta = 0.7$ and include terms up to the quadrupole order in the multiple expansion. The gas evolution is followed by means of smoothed particle hydrodynamics \citep[e.g.][]{Lucy1977}. A Plummer potential is used to soften gravitational forces, with a constant softening length $\epsilon = 200 \pc$. The galaxy models are evolved in {\it isolation} for $1 \Gyr$ before the start of merger simulation \citep{Chillingarianetal2010}.

Here, we consider a set of giant-dwarf minor merger models where the host galaxy is of Sa-type and the morphology of the satellite galaxy varies from dE0l to dSb.  Each minor merger model is referred as a unique string given by `{\sc [host galaxy][satellite galaxy][orbit ID][orbital spin]33}' where {\sc [host galaxy]} and {\sc [satellite galaxy]} denote their morphology types,  and {\sc [orbit ID]} denotes the orbit number as assigned in the GalMer library. {\sc [orbital spin]} denotes the orbital spin vector (`dir' for direct and `ret' for retrograde orbits), and `33' refers to $i_1 =33^{\circ}$. We follow this scheme of nomenclature throughout the paper. The key orbital parameters of the minor merger models considered here, are listed in Table.~\ref{table:key_param}. We define the epoch of merger, $T_{\rm merge}$, when the distance between the centre of mass of two galaxies becomes close to zero. The resulting $T_{\rm merge}$, along with the times of first and second pericentre passages for the selected minor merger models are also listed in Table.~\ref{table:key_param}.

\begin{table*}
\centering
\caption{Key parameters for the selected minor merger models from GalMer library}
\begin{tabular}{ccccccccccc}
\hline
model$^{(1)}$ & $r_{\rm ini}$$^{(2)}$ & $v_{\rm ini}$$^{(3)}$ & $L_{\rm ini}$$^{(4)}$ & $E_{\rm ini}$$^{(5)}$ &spin$^{(6)}$ &  Pericentre$^{(7)}$ & $T_{1,\rm  peri}$$^{(8)}$ & $T_{2, \rm peri}$$^{(9)}$ & $T_{\rm merger}$$^{(10)}$ & $T_{\rm end}$$^{(11)}$\\
& (kpc) & ($\times 10^2 \kms$) & ($\times 10^2 \kms \kpc$) & ($\times 10^4$ km$^2$ s$^{-2}$) &&  dist. (kpc) & (Gyr) & (Gyr) & (Gyr) & (Gyr)\\
\hline
gSadE001dir33 & 100 & 1.48 & 29.66 & 0. & up & 8.  & 0.5 & 1.1 & 1.55 & 3.8\\
gSadE001ret33 & 100 & 1.48 & 29.66 & 0. & down &8. &  0.5 & 1.3 & 1.95 & 3.8 \\
gSadE002dir33 & 100 & 1.52 & 29.69 & 0.05 & up & 8. & 0.45 & 1.2 & 1.55 & 3. \\
gSadE002ret33 & 100 & 1.52 & 29.69 & 0.05 & down & 8. & 0.45 & 1.4 & 2. & 3. \\
gSadE003dir33 & 100 & 1.55 & 29.72 & 0.1 & up & 8. & 0.45 & 1.25 & 1.95 & 3. \\
gSadE003ret33 & 100 & 1.55 & 29.72 & 0.1 & down & 8. & 0.45 & 1.5 & 2.25 & 3. \\
gSadE004dir33 & 100 & 1.48 & 36.33 & 0. & up & 8. & 0.5 & 1.2 & 1.7 & 3. \\
gSadE004ret33 & 100 & 1.48 & 36.33 & 0. & down & 8.  & 0.5 & 1.75 & 2.85 & 3. \\
gSadE006dir33 & 100 & 1.55 & 36.43 & 0.1 & up &  16. & 0.45 & 1.45 & 2. & 3. \\
gSadE006ret33 & 100 & 1.55 & 36.43 & 0.1 & down & 16. & 0.45 & 1.95 & 2.85 & 3. \\
gSadSb01dir33 & 100 & 1.48 & 29.66 & 0. & up & 8. & 0.45 & 1.1 & 1.85 & 3. \\
gSadSb01ret33 & 100 & 1.48 & 29.66 & 0. &  down & 8.& 0.45 & 1.35 & 2.85 & 3. \\
\hline
\end{tabular}
\centering
{ (1) GalMer minor merger model; (2) initial separation between two galaxies; (3)  absolute value of initial relative velocity; (4)  $L_{\rm ini} = |{\bf r}_{\rm ini} \times {\bf v}_{\rm ini}$|}; (5) $E_{\rm ini} = \frac{1}{2}v_{\rm ini}^2 -G(m_1+m_2)/r_{\rm ini}$, with $m_1 = 2.3 \times 10^{11} \Msun$, and  $m_2 = 2.3 \times 10^{10} \Msun$; (6) orbital spin; (7) pericentre distance; (8) epoch of first pericentre passage; (9)  epoch of second pericentre passage; (10) epoch of merger; (11) total simulation run time. Columns (2)-(7) are taken from \citet{Chillingarianetal2010}.
\label{table:key_param}
\end{table*}

\section{Evolution of stellar bars in GalMer models}
\label{sec:temp_bar}

\begin{figure}
\centering
\includegraphics[width=1.0\linewidth]{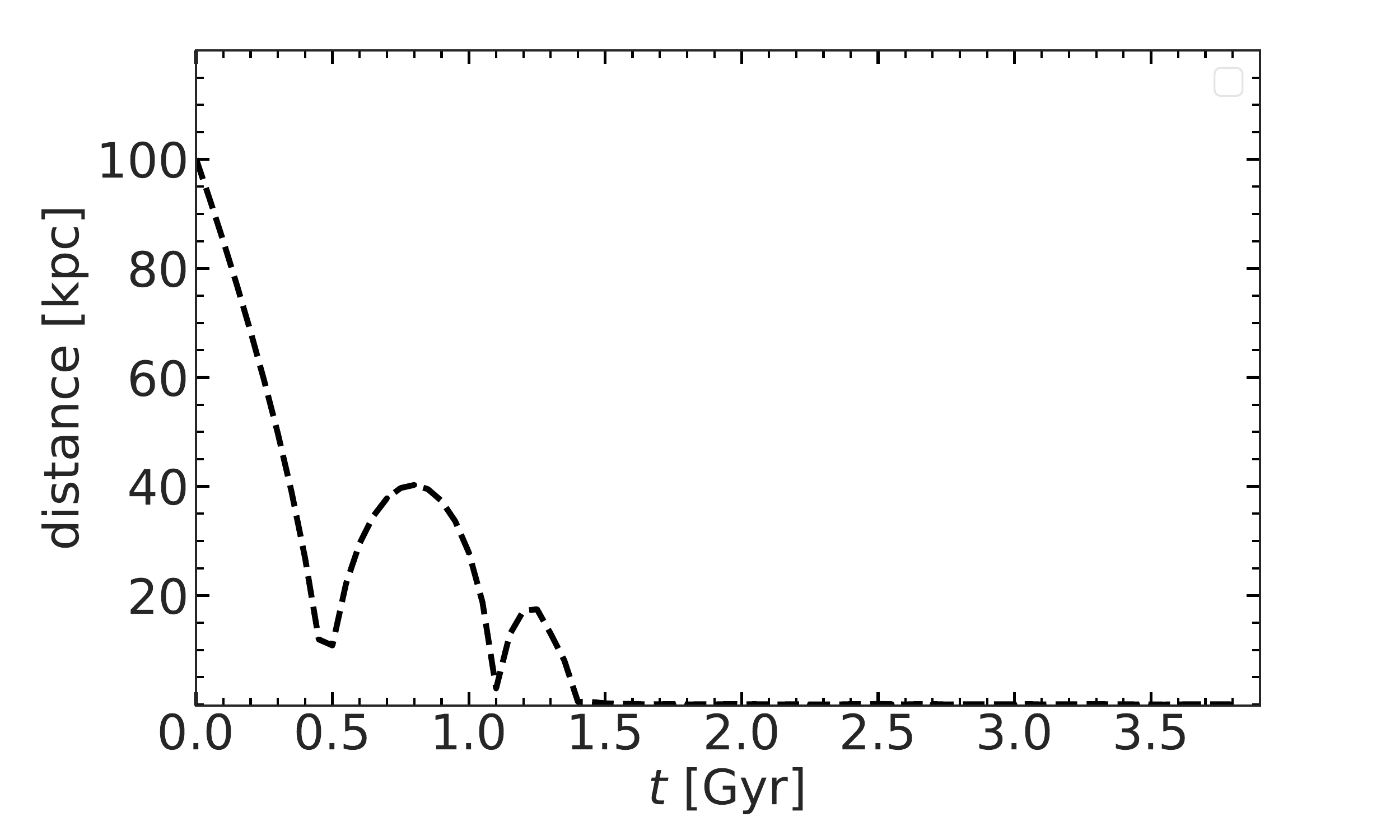}
\caption{ Distance between the centres of the satellite (dE0) and the host galaxy (gSa) shown as a function of time for the minor merger model {\bf gSadE001dir33}.}
\label{fig:dist_gsade001dir33}
\end{figure}

\begin{figure*}
\centering
\includegraphics[width=1.0\linewidth]{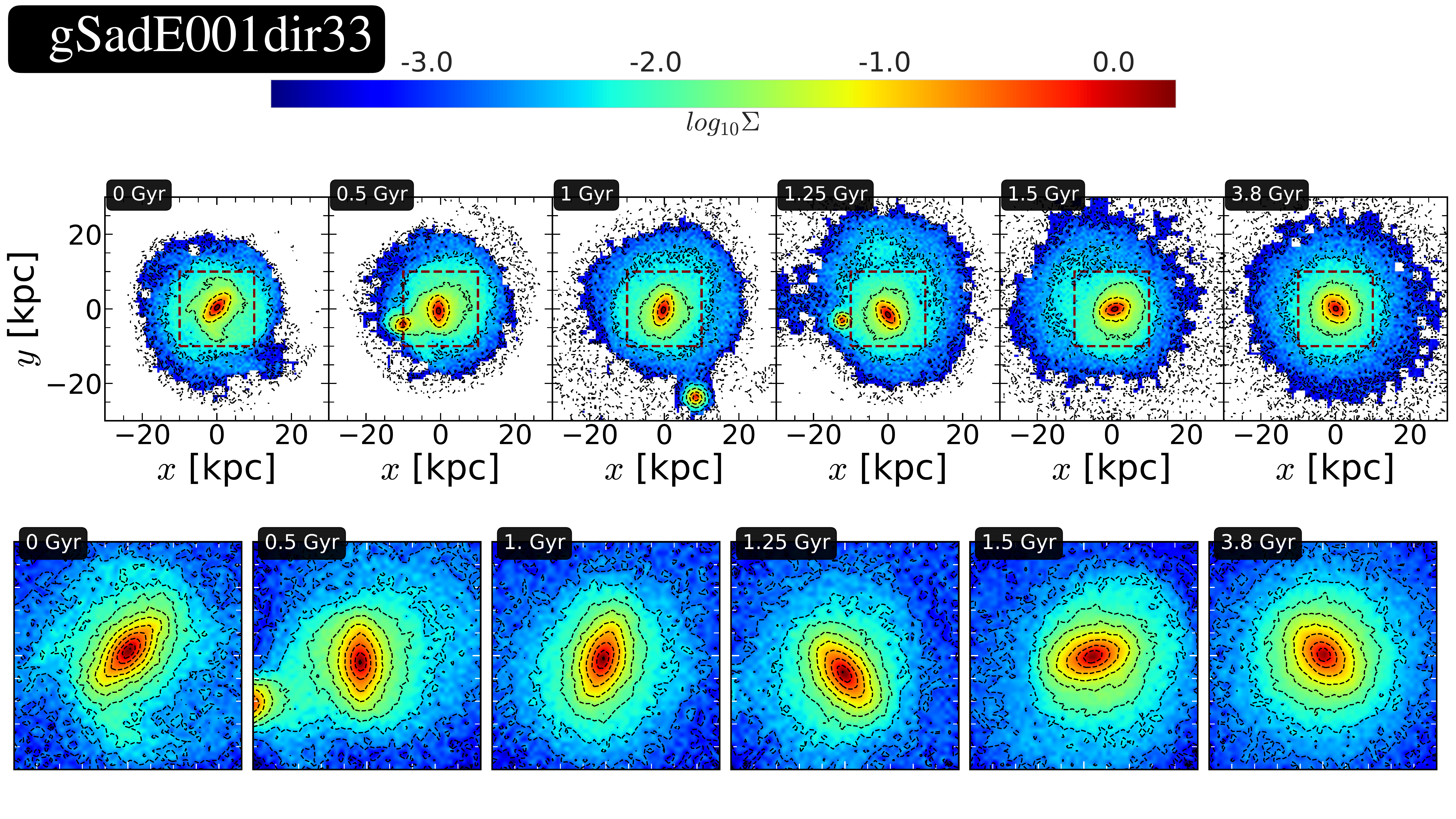}
\caption{{\it Top panels} : Face-on density distribution of host plus satellite (gSa+dE0) system, shown at different epochs for the minor merger model {\bf gSadE001dir33}. The rectangular boxes (in maroon) delineate the central $20 \kpc \times 20 \kpc$ region which includes the initial bar region. {\it Bottom panels}: zoom-in view of the central $20 \kpc \times 20 \kpc$ region. Black lines denote the contours of constant surface density.}
\label{fig:density_illustartion_collage}
\end{figure*}

Here, we investigate how a central stellar bar, initially present in a host galaxy, evolves after it suffers a minor merger (mass ratio 1:10) with a satellite galaxy. To do that, we choose a minor merger model {\bf gSadE001dir33} from the GalMer database where a dwarf E0-type galaxy merges with a host giant Sa-type galaxy. In the beginning, the host galaxy (gSa) harbours a prominent central stellar bar; thereby serving an ideal testbed for this work.  Fig.~\ref{fig:dist_gsade001dir33} shows the temporal evolution of distance between the centres of mass of these two galaxies. After each pericentre passage, the satellite loses a part of its orbital angular momentum due to the dynamical friction and the tidal torque. Consequently, it falls deep in the gravitational potential of the host galaxy and ultimately merges with the host galaxy. Fig.~\ref{fig:density_illustartion_collage} shows the face-on density distribution of stellar particles of the minor merger model {\bf gSadE001dir33} at six different epochs, before and after the merger. At the beginning ($t = 0 \Gyr$), the host (gSa) galaxy harbours a prominent stellar bar as delineated by the central elongated contours; however, at the end of the simulation ($t = 3.8 \Gyr$), the contours in the central region of the merger remnant are rounder in shape, suggestive of a bar weakening phenomenon.

To probe further, we created the face-on density maps at the beginning and at the end of the model {\bf gSadE001dir33}, and performed a multi-component decomposition of the radial density profiles. The radial profiles of surface density, the ellipticity ($\epsilon = 1-b/a$, $a$ and $b$ being semi-major and semi-minor axes, respectively) and the position angle (PA) are obtained by using {\sc {IRAF ellipse}} task. The extracted radial density profiles are then decomposed into disc+bulge or disc+bulge+bar (when the bar is present). The bulge is represented by a S\'{e}rsic profile with S\'{e}rsic index $n_{1}$, effective radius $R_{e,1}$ and effective surface density $I_{e,1}$. The disc is modelled with an exponential profile with central surface density $I_{d0}$ and disc scale-length $R_{\rm d}$. Additionally, when a bar is present, it is represented by another S\'{e}rsic profile with S\'{e}rsic index $n_{2}$, effective radius $R_{e,2}$, and effective surface density $I_{e,2}$ \citep{Elmegreenetal1996}.  Mathematically, all the components can be represented as:

\begin{equation}
\begin{split}
I(R) = I_{d0} e^{(-R/R_{\rm d})}+I_{e,1}e^{-b_{n_{1}}[(R/R_{e,1})^{(1/n_{1})}-1]}\\
+ I_{e,2}e^{-b_{n_{2}}[(R/R_{e,2})^{(1/n_{2})}-1]}\,,
\end{split}
\label{eq:photo_decomp_bar}
\end{equation}
\noindent where the multi-component fitting has been performed with {\sc PROFILER} software \citep{Ciambur2016}.

\begin{figure*}
\centering
\includegraphics[width=0.95\linewidth]{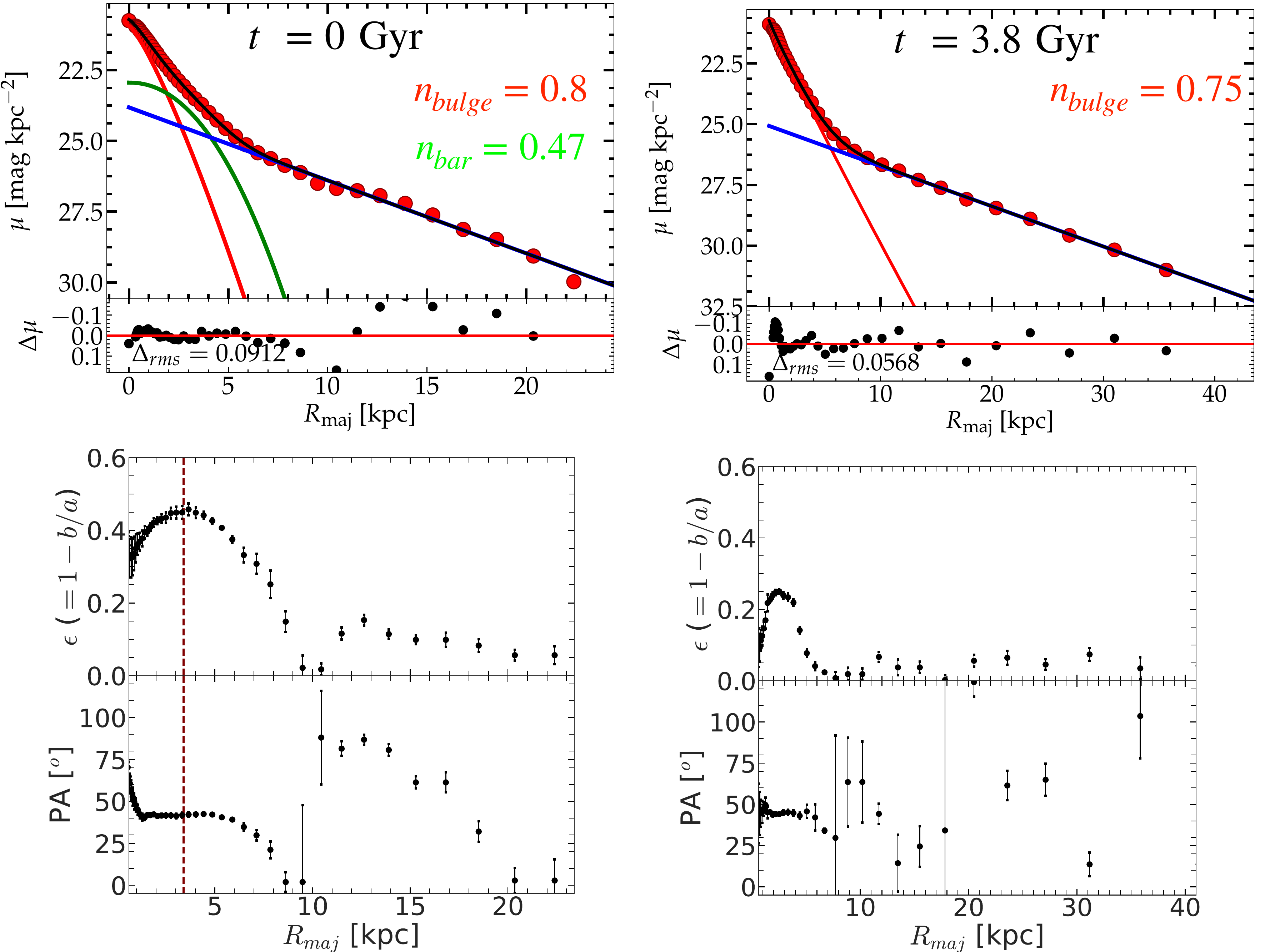}
\caption{{\it Top panels} show the multi-component decomposition of radial surface density profiles at the start ($t=0 \Gyr$) and the end of the simulation run ($t=3.8 \Gyr$) for the model {\bf gSadE001dir33}. Red and blue solid lines denote the S\'{e}rsic bulge and the exponential disc, respectively. The bar (when present) is denoted by another  S\'{e}rsic  profile (shown in green). {\it Bottom panels} show the corresponding radial profiles of ellipticity ($\epsilon$) and the position angle (PA). The vertical maroon line denotes the effective radius of the bar (when present).}
\label{fig:photometry_gsade001dir33}
\end{figure*}

Fig.~\ref{fig:photometry_gsade001dir33} shows the multi-component decomposition of radial surface density at the start ($t=0 \Gyr$) and the end of the simulation run ($t=3.8 \Gyr$), along with the corresponding radial profiles of the ellipticity ($\epsilon$) and the position angle (PA). The presence of a second S\'{e}rsic component (with $n_2$ = 0.47, and $R_{e,2} =2.53 \kpc$) at $t=0 $ clearly demonstrates the presence of  bar. This is further supported by a characteristic peak in the radial ellipticity ($\epsilon$) profile ($\epsilon_{max} \sim 0.36$) and constant position angle (PA) values in the central bar region. However, at the end of the simulation run ($t = 3.8 \Gyr$), the final morphology resembles an S0 galaxy with no discernible central bar (as indicated by the absence of a second S\'{e}rsic profile);  the peak value of the ellipticity ($\epsilon_{max}$) reduces to 0.14 at $t= 3.8 \Gyr$.

\begin{figure}
\centering
\includegraphics[width=1.0\linewidth]{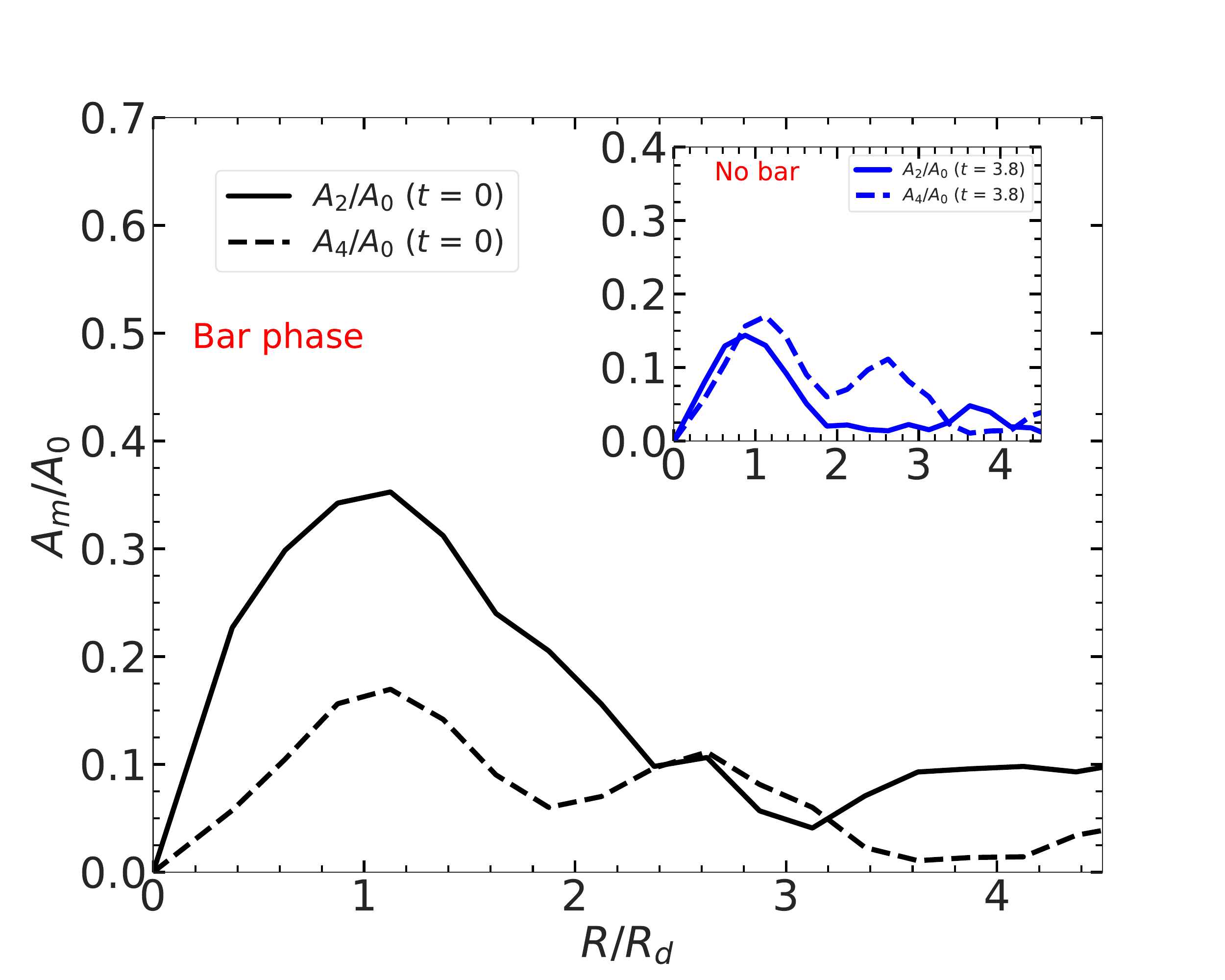}
\medskip
\includegraphics[width=1.0\linewidth]{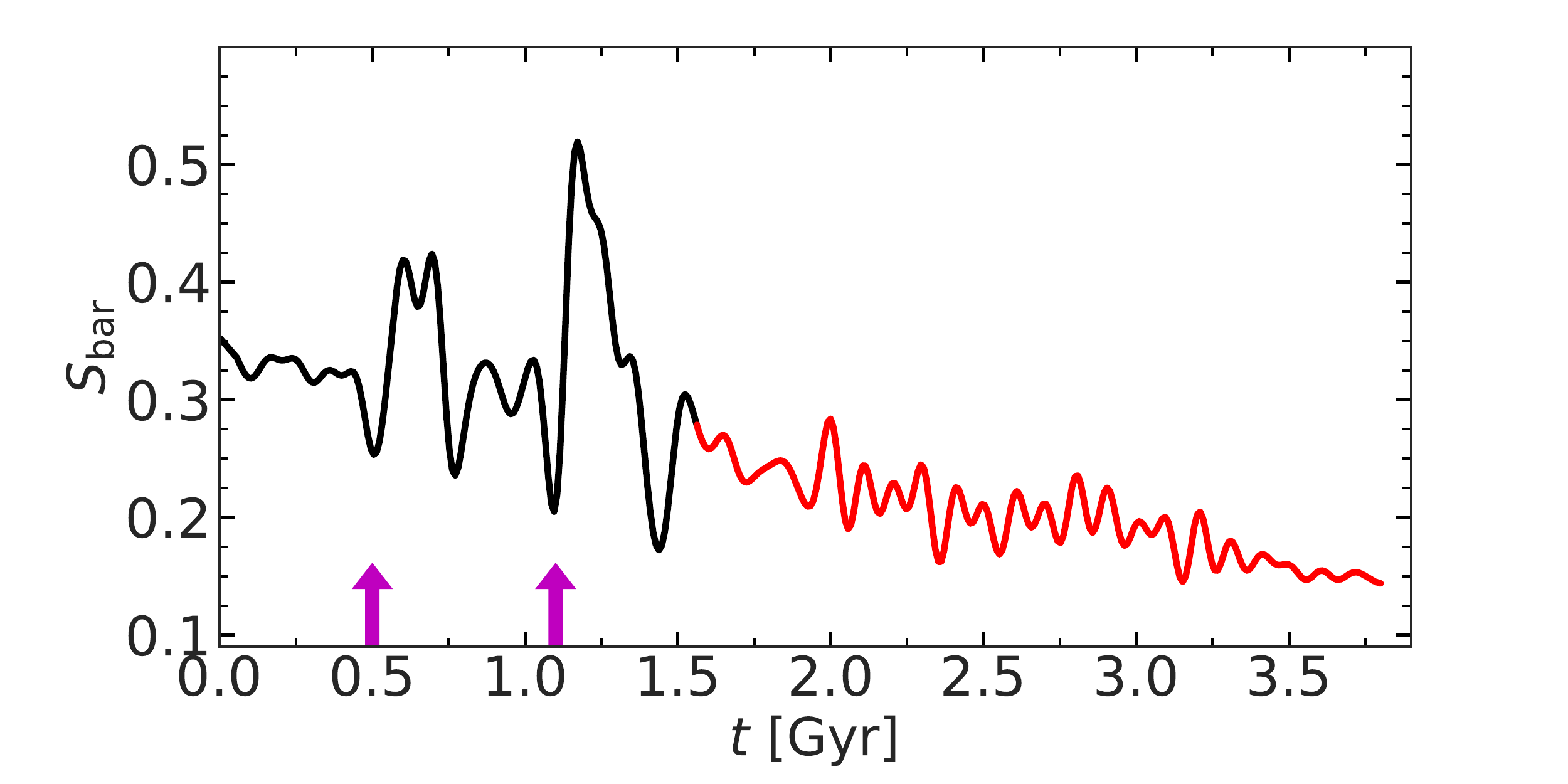}
\caption{{\it Top panel} shows the radial profiles of $m=2$ and $m=4$ Fourier coefficients  at the start ($t=0 \Gyr$) and the end of the simulation run ($t=3.8 \Gyr$, shown in inset) for the model {\bf gSadE001dir33}. The disc scale-length, $R_{\rm d}$ is $3\kpc$. {\it Bottom panel} shows the temporal evolution of the bar strength, $S_{\rm bar}$,  before (shown in black) and after (shown in red) the minor merger for the same model. The vertical arrows denote the epochs of first and second pericentre passages.}
\label{fig:fourier_gsade001dir33}
\end{figure}

To quantify the temporal change of the central stellar bar, we calculated the radial profiles of the amplitudes of the $m=2$ and $m=4$ Fourier modes ($A_2/A_0$ and $A_4/A_0$), at the beginning and at the end of the simulation {\bf gSadE001dir33}. This is shown in Fig.~\ref{fig:fourier_gsade001dir33} ({\it top panel}). At the beginning, the presence of the bar is clearly indicated by a peak (peak value $\sim 0.36$) in the radial profile of $m=2$ Fourier component in the central region; however, at $t = 3.8 \Gyr$, the peak value of the radial profile of $m=2$ Fourier coefficient $\sim 0.14$, indicating the bar has been substantially weakened. Also, at $t = 3.8 \Gyr$, the peak value of $m=4$ Fourier coefficient is small. We define the strength of the bar, $S_{\rm bar}$, at any given time $t$, as $S_{\rm bar} = (A_2/A_0)_{\rm max}$, where $A_m$ is the coefficient of $m{\rm th}$ Fourier harmonics.
Fig.~\ref{fig:fourier_gsade001dir33} ({\it bottom panel}) shows the corresponding temporal evolution of bar strength. After each pericentre passage, the initial bar gets stronger due to the tidal interaction, as indicated by the peaks in the $S_{\rm bar}$ (compare Figs~\ref{fig:fourier_gsade001dir33} and~\ref{fig:density_illustartion_collage}). This is in accordance with what has been shown previously where a bar instability can be excited in a tidal interaction \citep[e.g.,][and references in section~\ref{sec:intro}]{Peiranietal2009,Langetal2014,Lokasetal2014,Inmaetal2017}. However, after the satellite merges with the host galaxy, the bar strength decays steadily; at the end of the simulation ($t = 3.8 \Gyr$), the value of $S_{\rm bar}$ is $\sim 0.14$, thereby demonstrating the substantial weakening of the central stellar bar in the post-merger remnant of a minor merger.

\subsection{Comparison with isolated evolution}
\label{sec:isolated_evo}

\begin{figure}
\centering
\includegraphics[width=1.0\linewidth]{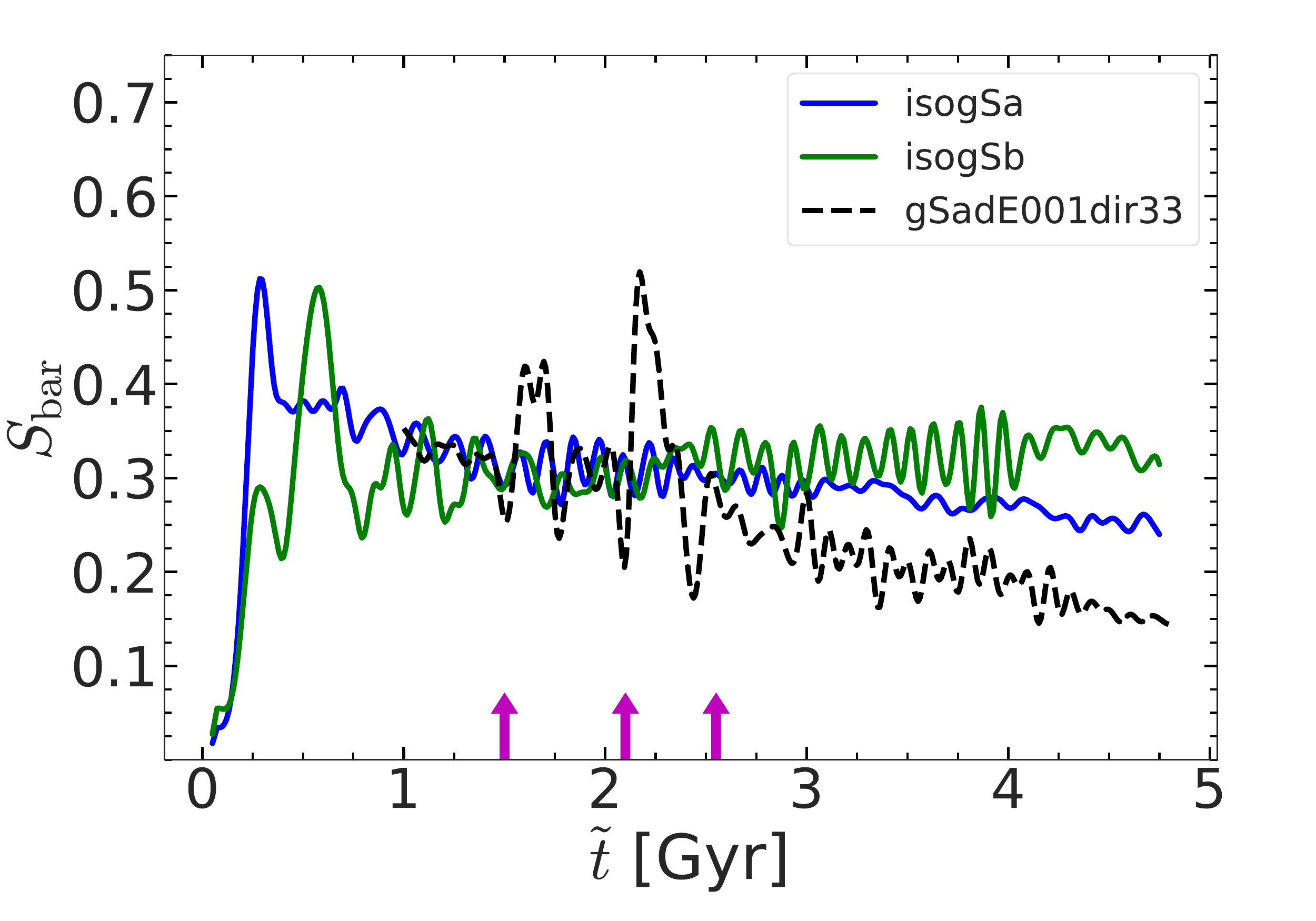}
\caption{\textbf{evolution in isolation :} temporal evolution of the bar strengths, $S_{\rm bar}$, calculated for the isolated models \textbf{isogSa} and \textbf{isogSb}, are compared with the same for the minor merger model \textbf{gSadE001dir33}. The vertical arrows bear the same meaning as in Fig.~\ref{fig:fourier_gsade001dir33}. Here, $\tilde t - t = 1 \Gyr$, for details see text.}
\label{fig:evolu_isolated}
\end{figure}

In the previous section, we have demonstrated that an initial bar gets weakened substantially in the post-merger remnant of the model \textbf{gSadE001dir33}. However, in order to attribute robustly the cause of the bar weakening to such a minor merger event, one needs to study the evolution of the bar strength in isolated models of the host galaxy. To achieve that, we have run the galaxy models of gSa- and gSb-type in isolation for $5 \Gyr$. The resulting temporal evolutions of the bar strength ($S_{\rm bar}$) for the isolated models \textbf{isogSa} and \textbf{isogSb} are shown in Fig.~\ref{fig:evolu_isolated}. We caution that, the minor merger simulation of \textbf{gSadE001dir33} starts from the time $\tilde t=1 \Gyr$ of the \textbf{isogSa} model (as also mentioned in section~\ref{sec:simu_setup}). Consequently, there is a time delay of $1 \Gyr$ between the isolated and the minor merger models, i.e., $\tilde t - t = 1 \Gyr$.

The formation of a strong bar starts around $1 \Gyr$ in both the isolated models. However, the isolated models are subjected to a secular evolution; the gas, present in the models, is driven inward, causing an increase in the angular momentum in the central region \citep[for details see][]{Minchevetal2012}. Consequently, the bar strength decreases in the following $1-1.5 \Gyr$ or so. From $\tilde t = 2.6-2.8 \Gyr$ onwards, the bar strength remains almost constant in the isolated models, whereas the bar strength in the minor merger model \textbf{gSadE001dir33} is seen to decrease monotonically except for a transient bar amplification phase after each pericentre passage of the satellite. This shows that the isolated models are able to host a stellar bar whose strength remains almost constant at later times. This clear difference in the temporal evolution of bar strength between the isolated and the minor merger model demonstrates that \textit{indeed} the minor merger event is liable for the substantial bar weakening in the post-merger remnant of a minor merger model.

\subsection{Dependence on the orbital parameters}
\label{sec:orb_dependence}

\begin{figure}
\centering
\includegraphics[width=1.0\linewidth]{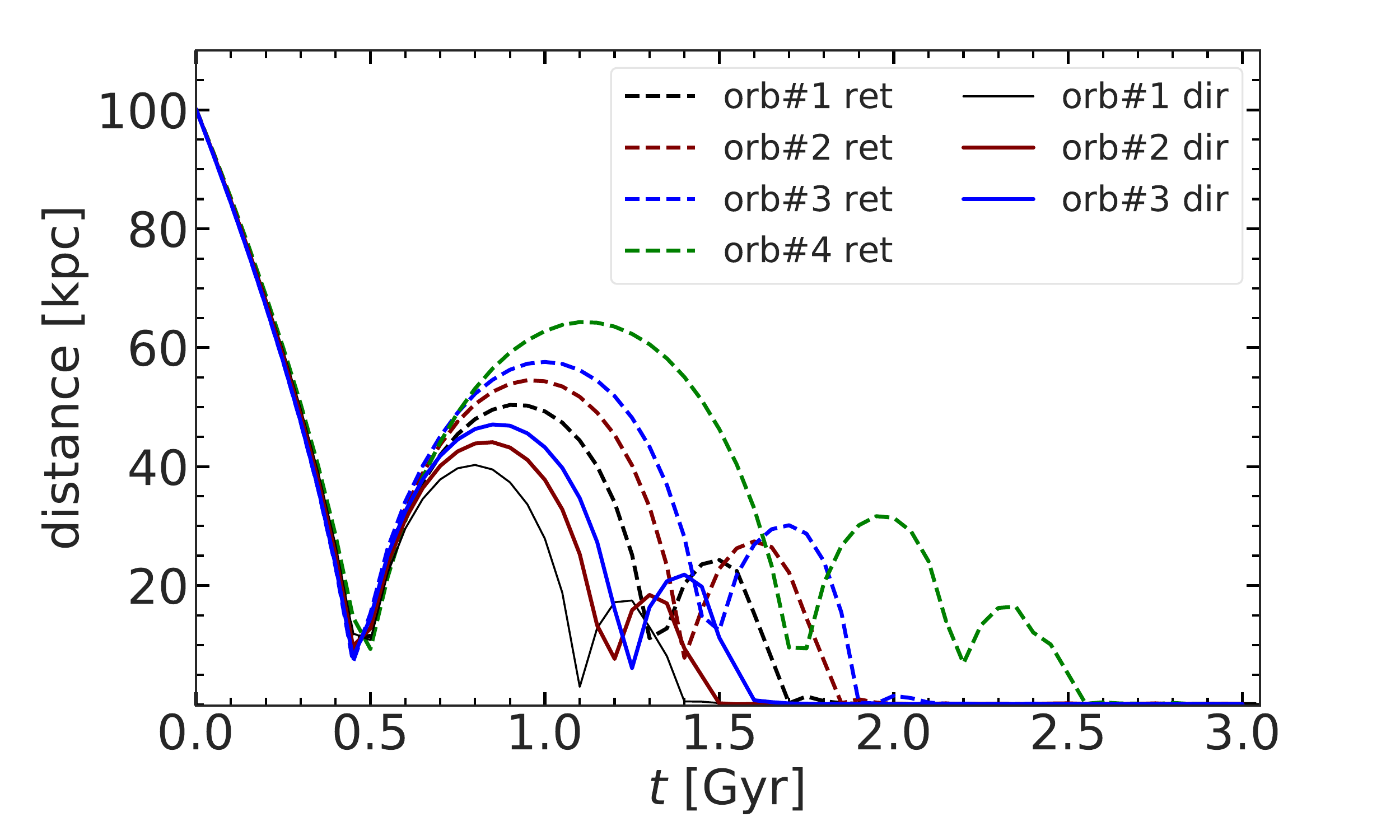}
\medskip
\includegraphics[width=1.0\linewidth]{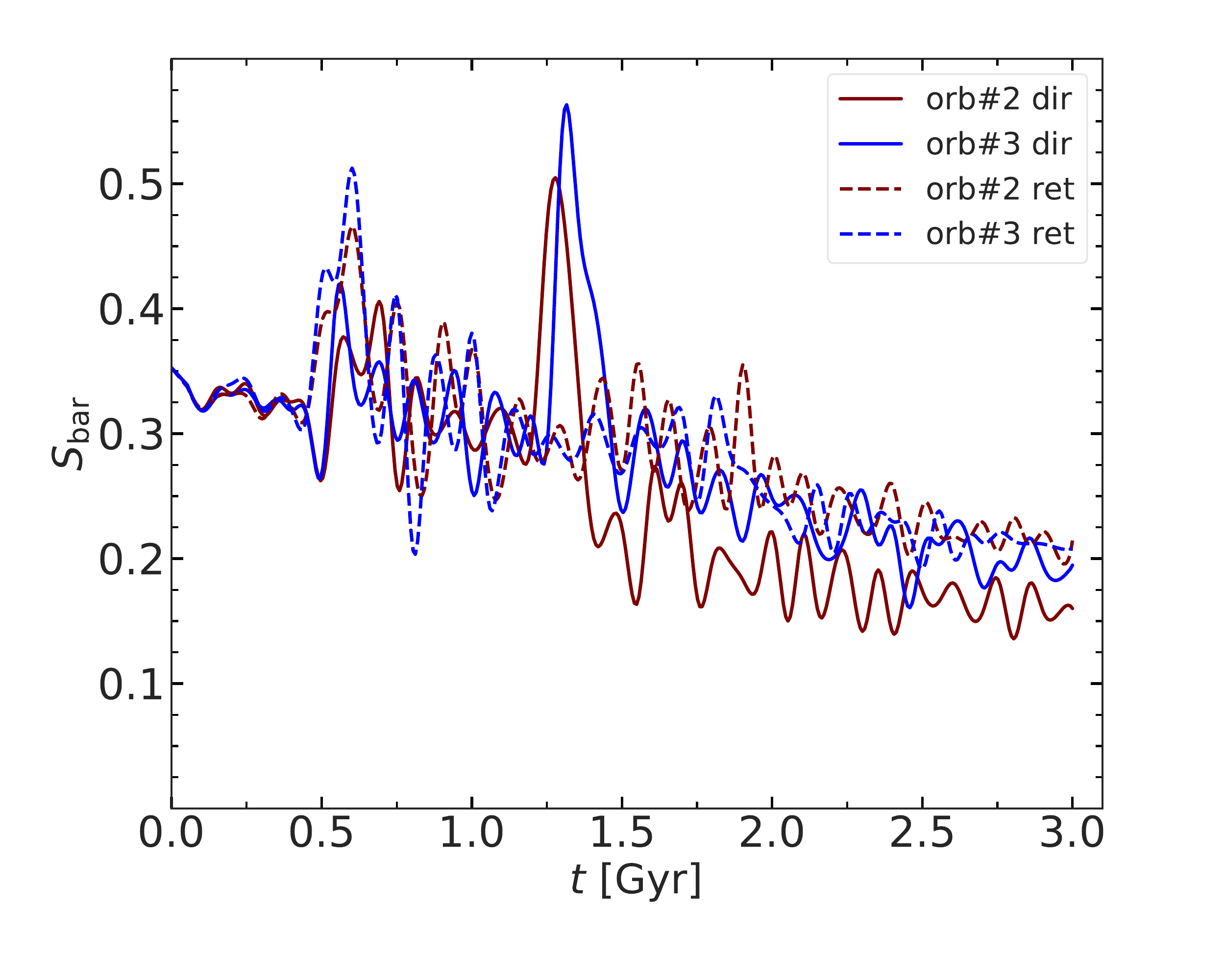}
\caption{{\it Top panel} shows the distance between the centres of the satellite (dE0) and the host galaxy (gSa) as a function of time, for different orbital configurations (for details see section~\ref{sec:orb_dependence}). {\it Bottom panel} shows the temporal evolution of the bar strength, $S_{\rm bar}$ for a few orbits with different orbital energies and orbital spin orientations (direct/retrograde). Solid and dashed lines denote direct and retrograde orbits, respectively.}
\label{fig:dist_orbparams}
\end{figure}

Here we explore different minor merger models with different orbital energies and orientation of the orbital spin vectors (direct/retrograde).
To do that, we choose minor merger models with higher orbit numbers. For these models, the orbital energies are higher than that for the {\bf gSadE001dir33} model (for details see section~\ref{sec:simu_setup}).  Fig.~\ref{fig:dist_orbparams} ({\it top panel}) shows the time evolution of distance between centres of mass of two galaxies for different orbital configurations considered here. We point out that the time of interaction, i.e., the time interval between the first pericentre passage and the time of merging ($T_{\rm merge}$), gets systematically enhanced as the orbital energy increases. This trend is much more pronounced when compared between a direct and a retrograde orbital configurations having the same orbital energy. 

We now investigate how the temporal evolution of the central bar in a minor merger scenario depends on the orbital energies and the orientation of the orbital spin vectors. First, we  performed a multi-component decomposition of the radial surface density profiles (as outlined in section~\ref{sec:temp_bar}) at the end of the simulation run for all minor merger models considered here. However, for the sake of brevity, these are not shown here. We noticed that, for these minor merger models, the resulting S\'{e}rsic `$n$' for bar is less than 0.47 (S\'{e}rsic `$n$' for bar obtained at $t=0$, see Fig.~\ref{fig:photometry_gsade001dir33}). This implies that the mass distribution of the central $m=2$ non-axisymmetric structure gets more flattened by the end of the simulation run. Also, the peak in the radial ellipticity profile ($\epsilon_{max}$) is diminished from its initial value ($\epsilon_{max} \sim 0.36$ at $t = 0 \Gyr$); thus, further supporting the fact that the central non-axisymmetric structure has become rounder by the end of the simulation run.

Next, we probe the dependence of the temporal evolution of the bar strength, $S_{\rm bar}$ on different orbital energies and orientation of the orbital spin vectors. This is shown in Fig.~\ref{fig:dist_orbparams} ({\it bottom panel}).We found that in all minor merger models considered here, each pericentre passage of the satellite produces a transient increase in the bar strength ($S_{\rm bar}$); a scenario similar to the case of  {\bf gSadE001dir33} model. This finding is at par with the past studies which demonstrated the triggering of bar mode in disc galaxies due to the tidal encounter \citep[e.g.,][]{Peiranietal2009,Langetal2014,Lokasetal2014,Inmaetal2017}. However, once the satellite merges with the host galaxy, the bar strength decreases steadily. At the end of the simulation run ($t = 3 \Gyr$), the $S_{\rm bar}$ values range in $0.18-0.22$, denoting a substantial bar weakening phenomenon, similar to the trend seen for the model {\bf gSadE001dir33}. The steady decreasing trend of $S_{\rm bar}$ implies that if these models were evolved for another $0.5-1 \Gyr$, the $S_{\rm bar}$ values in the post-merger remnants would have reached to $\sim 0.14$ or so, similar to what is seen in the {\bf gSadE001dir33} model. This shows that the bar weakening phenomenon in minor merger events is a generic process, irrespective of the orbital energies and the orientation of the orbital spin vector.

\section{Physical causes of bar weakening}
\label{sec:physics_barweakening}

In the previous sections, we demonstrated that a minor merger leads to a steady decrease in the bar strength implying the bar weakening event. This trend holds true for different orbital parameters (e.g., orbital energy, orientation of orbital spin vector). Here we explore the underlying physical mechanisms which are liable for the bar weakening.

\subsection{Central mass enhancement}
\label{sec:central_massincrease}

Past theoretical studies showed that a massive central mass concentration can destroy/weaken a stellar bar. However, this process might require a very high central mass content \citep[$\sim 5$ per cent of the disc mass, see e.g., ][]{ShenSellwood2004,Athanassoulaetal2005,HozumiHernqusit2005}. Here, we investigate how the mass concentration in the central region (encompassing the bar) changes with time, before and after the minor merger occurs. 

\begin{figure}
\centering
\includegraphics[width=\linewidth]{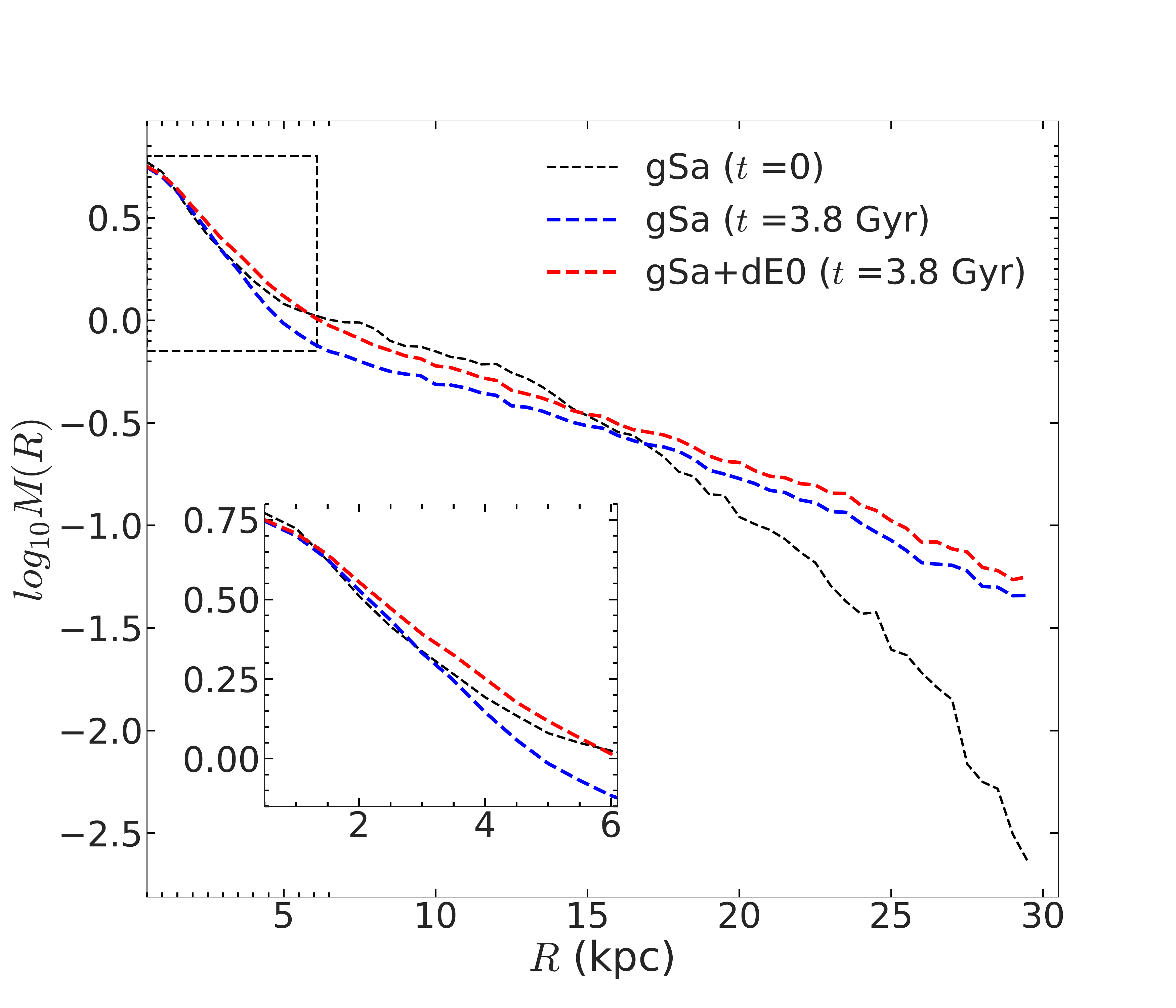}
\caption{Radial mass distribution, calculated at the beginning and at the end ($t=3.8 \Gyr$) of the simulation run for the model {\bf gSadE001dir33}. Blue dashed line denotes the contribution from the host (gSa) galaxy whereas red dashed line shows for the host plus satellite (gSa+dE0) system. The central bar region ($R \leq 6.1 \kpc$ as indicated within the dashed black box) is shown in the inset. The mass is in units of $2.25 \times 10^9 \Msun$}
\label{fig:mass_increase_individual}
\end{figure}

Fig.~\ref{fig:mass_increase_individual} ({\it top panel}) shows the radial mass distribution at the beginning and at the end ($t = 3.8 \Gyr$) of the model {\bf gSadE001dir33}. We calculated the radial mass profiles, first considering only the stellar particles from the host galaxy, and then, taking all the stellar particles from both the host and the satellite galaxy. This scheme, in turn, will reveal the relative contribution of the host and the satellite galaxy separately in the net mass change within the central bar region. Fig.~\ref{fig:mass_increase_individual} reveals that the initial, centrally-concentrated stellar particles of the host galaxy is dispersed at larger radii at later epochs. This in turn, leads to a decrease in the total mass of the central bar region, and a flattened mass distribution at larger radii from the centre. However, when the stellar particles of both the host and satellite galaxies are considered, the radial mass distribution, at $t = 3.8 \Gyr$, displays a net mass enhancement in the central bar region. This shows the accumulation of stellar particles from the satellite galaxy is liable for the net mass increment within the bar region. 

\begin{figure}
\centering
\includegraphics[width=1.05\linewidth]{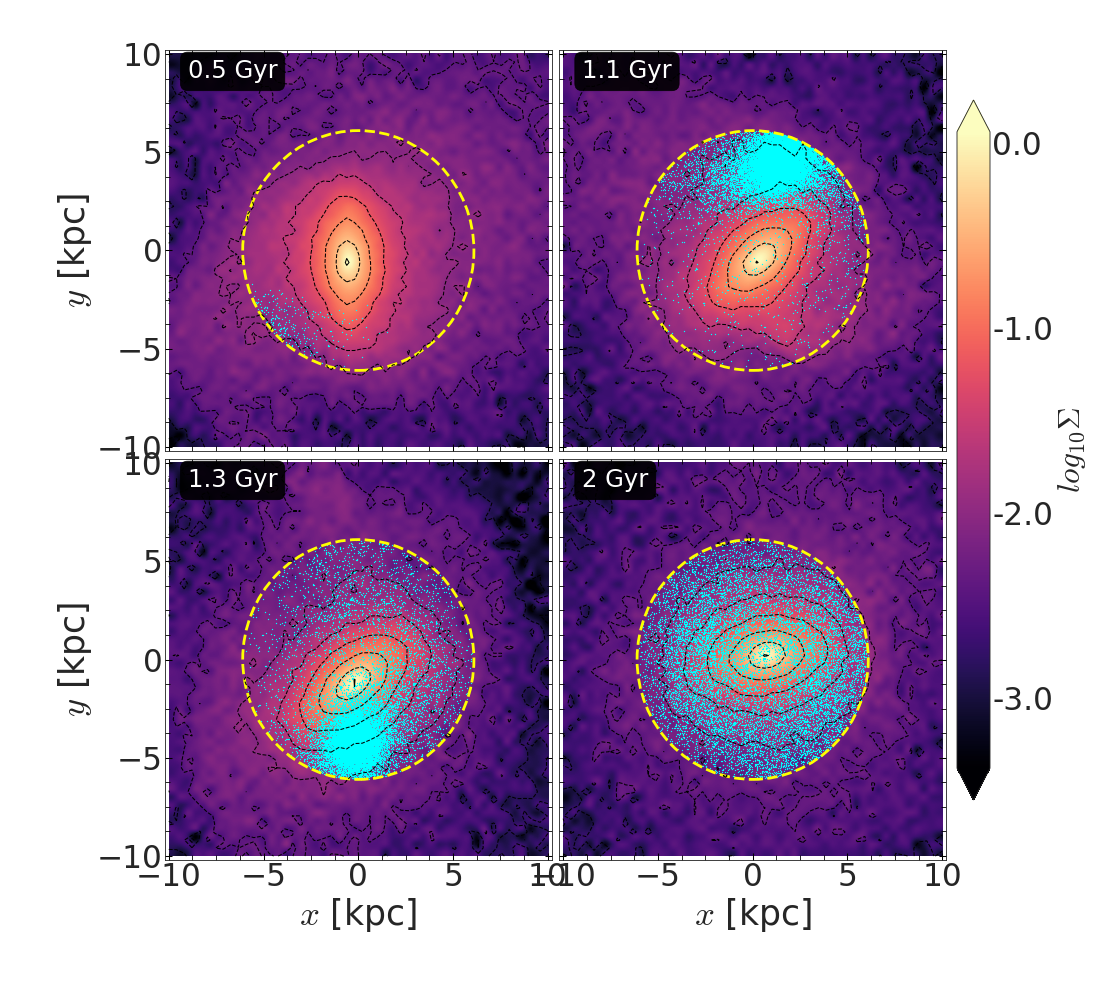}
\medskip
\includegraphics[width=1.05\linewidth]{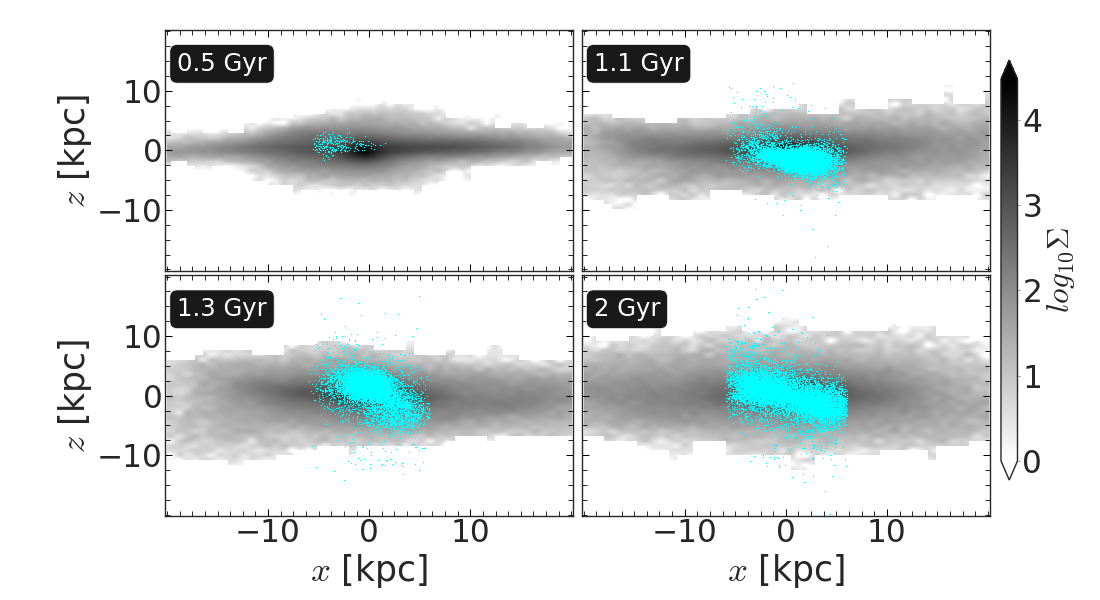}
\caption{ Face-on ({\it top panels}) and edge-on ({\it bottom panels}) density maps of the host galaxy shown at four different epochs of the model {\bf gSadE001dir33}. The stellar particles from the satellite (cyan dots) accumulated within the initial bar region are over-plotted. The dashed circle (in yellow) in the top panels indicated the initial bar radius ($R_{\rm bar} = 6.1 \kpc$).}
\label{fig:mass_increase_denmaps}
\end{figure}

Next, we probe in details the distribution of accumulated stellar particles from the satellite in the central bar region. Fig.~\ref{fig:mass_increase_denmaps} further demonstrates the steady accumulation of satellite's particles within the central bar region. We point out that, the accumulated stellar particles from the satellite are {\it not} aligned preferentially in the disc plane; rather they are distributed over the whole bar region, and the distribution is vertically extended. Even for a direct orbital configuration, the angle of inclination is not zero ($i_1 = 33 \degrees$, see section~\ref{sec:simu_setup}), which in turn prevents the accumulated particles to be aligned in the disc plane. These accumulated stellar particles from the satellite participates in forming a thick-disc in the post-merger remnant \citep{Quetal2011a,Quetal2011b}. The physical implication of this accumulation process in context of the bar weakening process is discussed later.

\begin{figure*}
\centering
\includegraphics[width=0.9\linewidth]{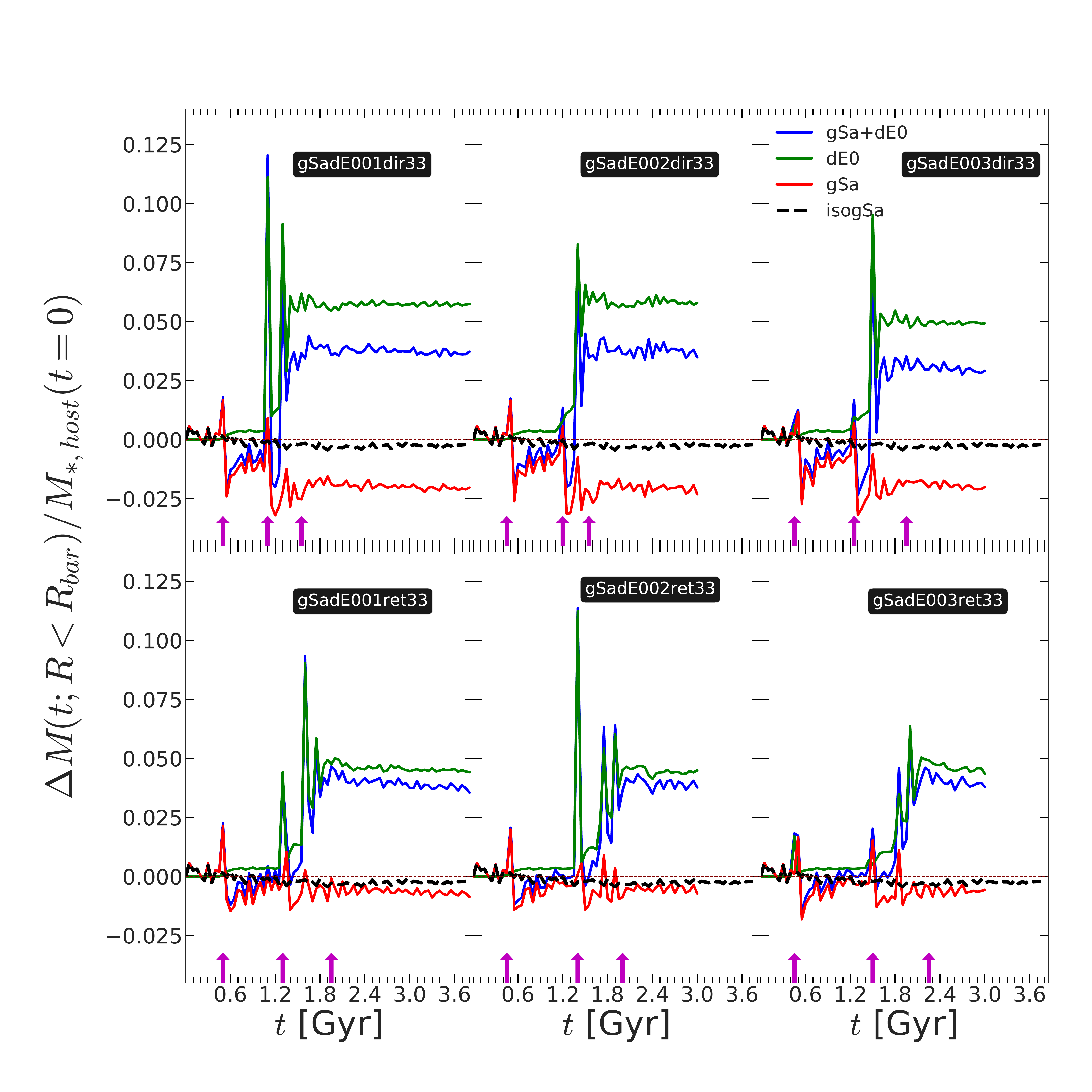}
\caption{Change of mass, $\Delta M (t; R < R_{\rm bar})$, averaged within the initial bar region ($R \leq 6.1 \kpc$), shown as a function of time, for different models. The averaging is done while taking the stellar particles from host (gSa), satellite (dE0), and host plus satellite galaxies (red, green, and blue lines), respectively. The change of mass is normalised by the total stellar mass of the host galaxy at $t=0$. Vertical arrows (in magenta) indicate the epochs of first and second pericentre passages and the epoch of merger, respectively. Black dashed line denotes the same for the isolated evolution of the host galaxy (\textbf{isogSa}).}
\label{fig:temporal_mass_increment}
\end{figure*}

Finally, we calculated the temporal evolution of the mass increase within the bar region. We measure that, at $t=0$, the central bar extent ($R_{\rm bar}$) is $\sim 6.1 \kpc$. For uniform comparison, we kept the extent of the bar region fixed at $6.1 \kpc$ at later time-steps. We checked that the bar extent varies less than 20 per cent of its initial extent during the entire `bar phase', and therefore would not introduce any artefact in the subsequent analyses. Fig.~\ref{fig:temporal_mass_increment} shows the corresponding temporal evolution of the change in the central mass content within the bar region ($\Delta M_{\rm net} (t; R< R_{\rm bar})$, defined in Eq.\ref{eq:dM}), for different models considered here. The net change in the mass within the bar region for the host plus satellite system is given as

\begin{equation}
\Delta M_{\rm net} (t; R< R_{\rm bar}) = \Delta M_{\rm host} (t; R< R_{\rm bar})+\Delta M_{\rm sat} (t; R< R_{\rm bar})\,,
\label{eq:dM}
\end{equation}
\noindent where $\Delta M_{\rm host}$, and $\Delta M_{\rm sat}$ are the contributions from the host and the satellite galaxy, respectively; they are calculated as
\begin{equation}
\Delta M (t; R< R_{\rm bar})  = M (t; R< R_{\rm bar})-M(t = 0; R< R_{\rm bar})\,.
\end{equation}

Fig.~\ref{fig:temporal_mass_increment} brings out the fact that $\Delta M_{\rm host}$, and $\Delta M_{\rm sat}$ have  an opposite effect within the bar region. The mass fraction of the stellar particles of host galaxy decreases with respect to the initial epoch. However, after the merger happens, a fraction of stellar particles of the satellite galaxy gets trapped within the bar region (as also shown in Fig.~\ref{fig:mass_increase_denmaps}). Hence, the net change in mass fraction within the bar region will be determined by the dominant of these two opposite effects of the host and the satellite galaxies. Fig.~\ref{fig:temporal_mass_increment} shows the general trend that, within the bar region, the mass accumulation from the satellite galaxy always dominates over the instantaneous mass loss from the host galaxy. This trend holds for the dE0-type satellite galaxy, and for different orbital energies and the orientation of the orbital spin vectors considered here.

Now, we briefly compare the bar weakening by mass accumulation scenario during the minor mergers with the past literature of bar weakening via growth of central mass concentration (hereafter CMC). We note that, the mass accumulation within the central bar region varies from $3.1 \times 10 ^9 \Msun$ to $4.6 \times 10^9 \Msun$ (equivalently, $\sim 3-4$ per cent of the total stellar mass of the host galaxy) for the direct orbits.  Similarly, for retrograde orbits, the  the mass accumulation within the central bar region varies from $4 \times 10 ^9 \Msun$ to $5.75 \times 10^9 \Msun$ (equivalently, $\sim 3.5-5$ per cent of the total stellar mass of the host galaxy). Past studies on bar destruction/weakening by CMC quoted the required mass to be $\sim 5$ per cent of the disc mass \citep[e.g., see][]{ShenSellwood2004,Athanassoulaetal2005,HozumiHernqusit2005}. Therefore, the amount of mass accumulation seen in our minor merger models is within the estimated range as reported in the past literature. However, we point out that it is important up to what extent the central mass is being accumulated. In the past studies of bar destruction due to enhancement in the CMC, the typical extent of CMCs is a few hundred parsecs to $2 \kpc$ or so \citep{ShenSellwood2004,HozumiHernqusit2005} which is consistent with the sizes of the molecular gas concentration \citep[e.g.,][]{Sakomotoetal1999,Reganetal2001}. On the other hand, the increment in the central mass concentration is measured within the initial bar radius ($\sim 6.1 \kpc$).

Also, we point out that in past numerical simulations, the CMCs were introduced in an adiabatic fashion. The time for full growth of CMC ($t_{\rm grow}$) varies from 0.7-1.5 Gyr \citep[e.g., see][]{ShenSellwood2004,Athanassoulaetal2005,HozumiHernqusit2005}. Thus, the (simulated) galaxy could readjust itself to the secular change of the underlying potential.
On the other hand, in our selected minor merger models, the mass accumulation in the central bar region happens in a rather short time-span after the merger happens ($\sim150-250  \Myr$). Thus, the merger remnant could not readjust itself to the abrupt change in the underlying potential.

Lastly, we comment about any possible effect of the `hybrid particle' scheme for representing the interstellar gas on the estimates of the central mass accumulation associated with the bar weakening process. We point out that for a longer depletion time, the hybrid particles do not follow the `exact'  hydrodynamical and the gravitational evolutions \citep[for details see][]{MihosandHernquist1994}.  We measured the temporal evolution of the change of mass of these hybrid particles within the bar region, and found that, the change in the mass concentration of these hybrid particles are only $\sim 1$ per cent of the total stellar mass of the host galaxy. The temporal evolution of the gas fraction of these hybrid particles within the bar region, and their plausible impact on the main findings of this paper are discussed further in details in section~\ref{sec:few_points}.

\subsection{Angular momentum exchange}
\label{sec:angmom_variation}

Past studies have demonstrated that a bar can grow in amplitudes by transferring the disc angular momentum to the dark matter halo;  this transfer takes place at the bar resonances \citep[e.g., see][]{TremaineWeinberg1984,HernquistWeinberg1992,DebattistaSellwood2000,Athanassoula2002,SellwoodandDebattista2006,Dubinskietal2009,SahaNaab2013}. On the other hand, \citet{Bournaudetal2005} showed that the angular momentum transfer, from the gas inflow to the stellar bar, can potentially weaken the bar. Here, we study in detail the angular momentum exchange in a minor merger event. During a minor merger event, the orbital angular momentum is distributed in both the host and the satellite galaxies where the satellite always gains a part of the orbital angular momentum, irrespective of orbital energy, and orientation \citep[for details see][]{Quetal2010}. The detailed redistribution of internal angular momentum into different components, namely, disc, bulge, and dark matter (hereafter DM) halo of the host galaxy in shown in Appendix~\ref{appen:int_angmom}.

Here, we focus on the central bar region ($R < R_{\rm bar}$) and study in detail the temporal evolution of angular momentum (hereafter AM) within the bar region during the bar weakening process. As in section~\ref{sec:central_massincrease}, we keep the extent of the bar region fixed at $6.1 \kpc$ while measuring the change in the specific AM. 
At time $t$, we calculate the $z$-component of the specific AM of the stellar particles of the host or the satellite galaxy, within the bar region using the definition
\begin{equation}
l_z (t; R < R_{\rm bar}) = \frac{1}{N(t)}\sum_{i=1}^{N(t)} \left[x_i(t) v_{y_i}(t)- y_i(t) v_{x_i}(t)\right]\,,
\end{equation}
\noindent where $N(t)$ is the total number of stellar particles contained within the bar region at time $t$, and $x, y, v_x, v_y$ are the position and velocity of the particles. The corresponding change in the internal specific AM is calculated as 

\begin{equation}
\Delta l_z (t; R < R_{\rm bar})  = l_z (t; R < R_{\rm bar})-l_z(t = 0; R < R_{\rm bar})\,.
\end{equation}
\noindent However, the change in the specific AM within the bar region due to the {\it host plus satellite} system is calculated, using the definition of total differential, as

\begin{equation}
\begin{split}
\Delta l_{z, net}(t; R < R_{\rm bar}) = l_{z,net}(t; R < R_{\rm bar}) \times\\
 \left[\frac{\Delta L_{z,net}(t; R < R_{\rm bar})}{L_{z,net}(t)}- \frac{\Delta M_{net}(t; R < R_{\rm bar})}{M_{net}(t; R < R_{\rm bar})}\right]\\
\end{split}
\label{eq:specangmom_change}
\end{equation} 

\noindent where $L_z(t; R < R_{\rm bar})$ is the $z$-component of the AM within the bar region at time $t$. The subscript `net' denotes the quantities that are calculated by taking into account the stellar particles from both the host and the satellite galaxies within the bar region. We caution that for calculating $\Delta l_{z, net}(t; R < R_{\rm bar}) $, the individual change of specific AM from the host and the satellite galaxies  {\it can not }be simply co-added (unlike the case of $\Delta M_{\rm net} (t; R< R_{\rm bar}) $), and has to be calculated using Eq.~\ref{eq:specangmom_change}. 

\begin{figure*}
\centering
\includegraphics[width=\linewidth]{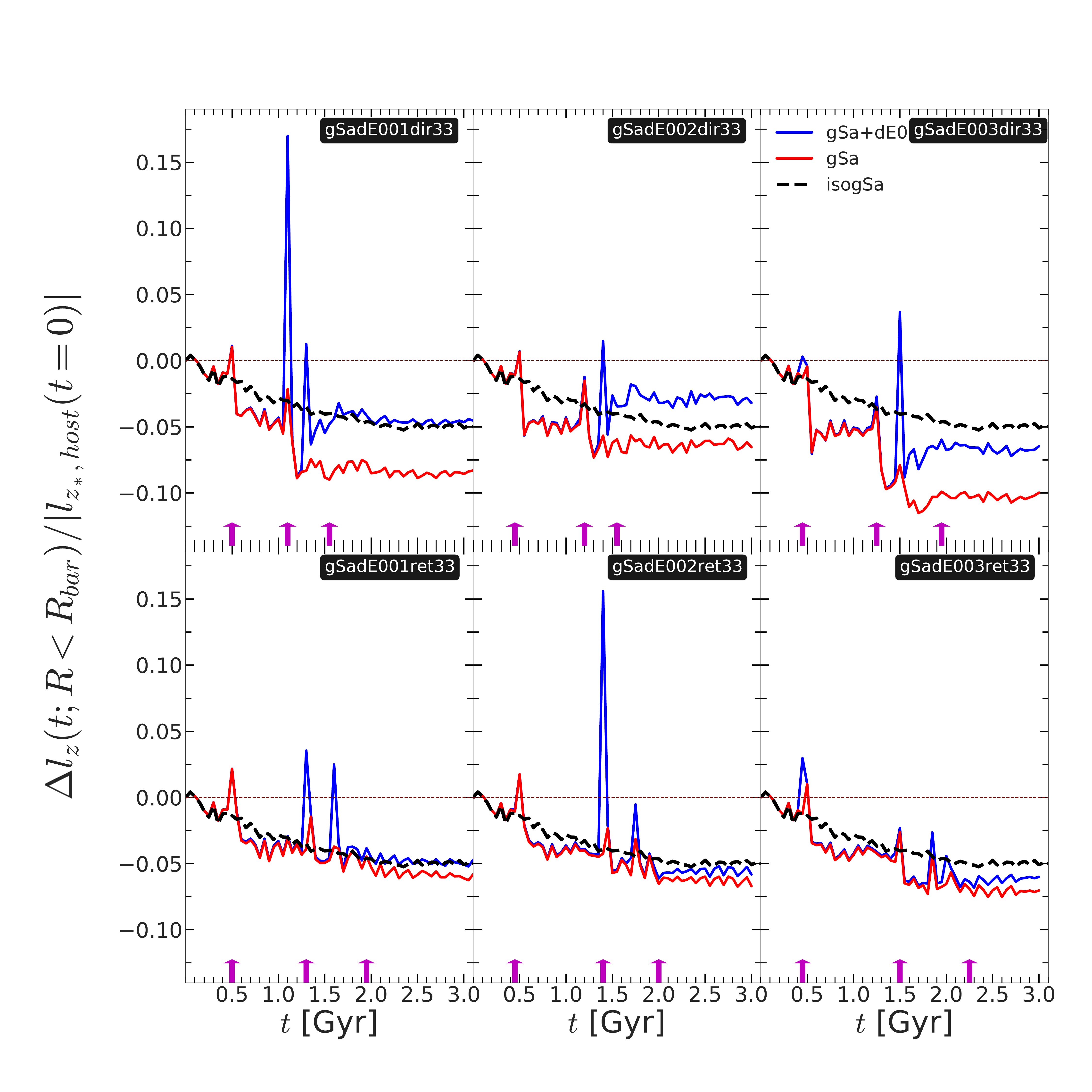}
\caption{Change of $z$-component of specific angular momentum, $\Delta l_z (t; R < R_{\rm bar})$, averaged within the initial bar region ($R \leq 6.1 \kpc$), shown as a function of time, for different models. The averaging is done while taking the stellar particles from host (gSa), and host plus satellite galaxies (red and blue lines), respectively. 
$\Delta l_z (t; R < R_{\rm bar})$ is normalised by the $z$-component of specific angular momentum of all stellar particles of the host galaxy at $t =0$. Vertical arrows (in magenta) indicate the epochs of first and second pericentre passages and the epoch of merger, respectively. Black dashed line denotes the same for the isolated evolution of the host galaxy (\textbf{isogSa}).}
\label{fig:temporal_specangmom_increment}
\end{figure*}

Fig.~\ref{fig:temporal_specangmom_increment} shows the temporal evolution of the change in the specific AM content within the bar region, for different minor merger models. The specific AM of the host galaxy within the bar region decreases with time, and this holds true for both the direct and retrograde orbital configurations. However, the specific AM loss within the bar region due to the host galaxy alone is more for a direct orbit than for a retrograde orbit with same orbital energy (compare top and bottom panels of Fig.~\ref{fig:temporal_specangmom_increment}). 
We checked that, the stellar particles from the satellite galaxy which eventually get trapped within the central bar region, contain high specific AM; thereby bringing in specific AM within the bar region. The net loss of specific AM within the bar region for the host plus satellite system is thus less when compared to the specific AM loss from the host galaxy alone. In other words, some fraction of specific AM loss due to the host galaxy is compensated by the fresh addition of stellar particles from the satellite galaxy having high specific AM.

\begin{figure}
\centering
\includegraphics[width=1.05\linewidth]{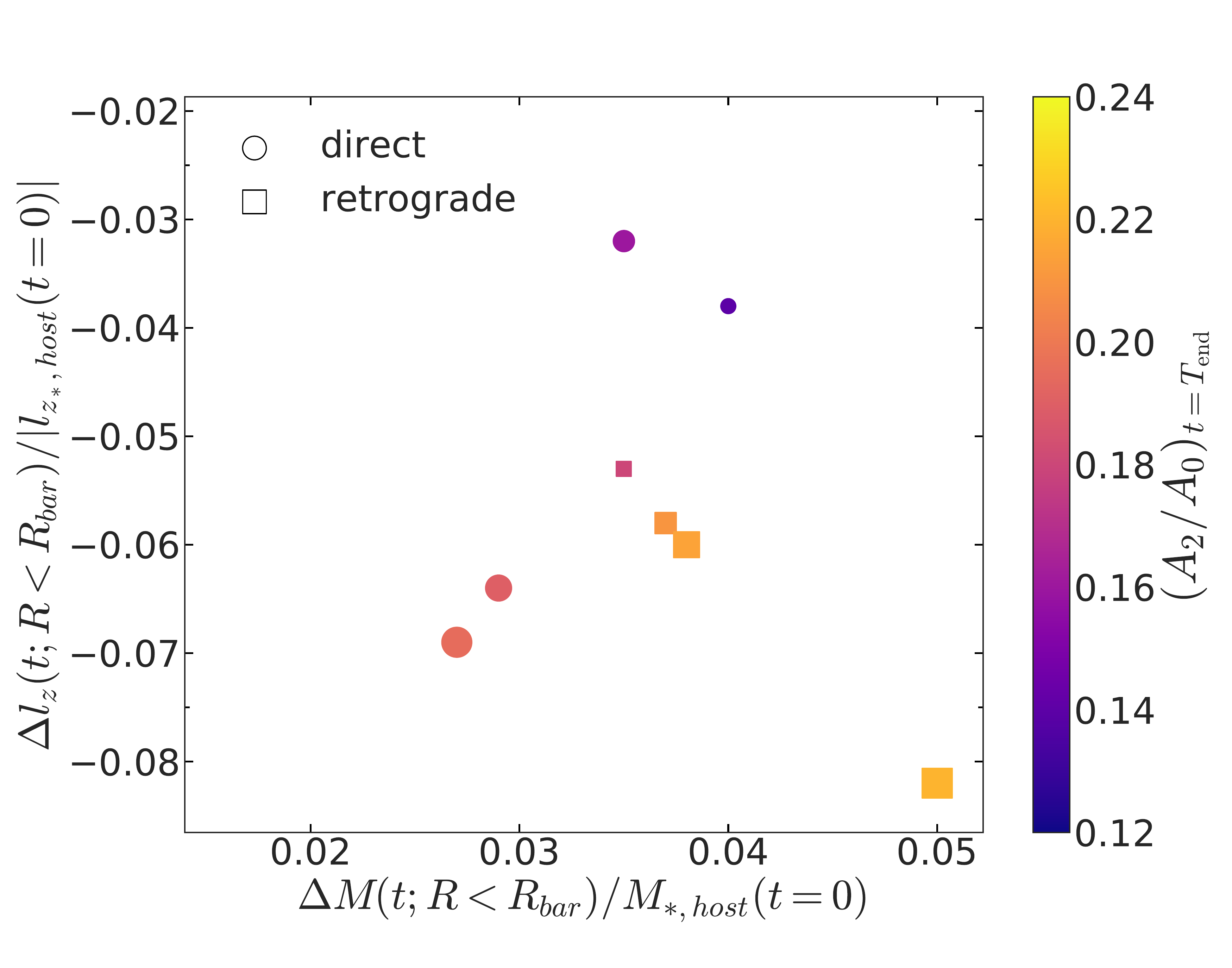}
\caption{Distribution of temporal change of mass and specific AM within the bar region is shown for the minor merger models considered here. The points are colour-coded by the peak value of $(A_2/A_0)$, calculated at the end the simulation run ($T_{\rm end}$). Circles and squares denote the direct and the retrograde orbits, respectively. The increasing size of the points denote higher orbit number (for details see section~\ref{sec:simu_setup}).}
\label{fig:delMdellz_combined}
\end{figure}

Finally, Fig.~\ref{fig:delMdellz_combined} compares the joint effects of temporal mass change and the specific AM change within the bar region, for different minor merger models. For the direct orbits with increasing orbital energy, the mass increase is progressively less, as reflected in the lesser values of $\Delta M_{\rm net} (t; R< R_{\rm bar}) $. Also, the loss in specific AM within the bar region is progressively more with increasing orbital energy. These together cause progressively lesser degree of bar weakening. However, for retrograde orbits, the trend is seen to differ from the direct orbits. For retrograde orbits, both the mass increase and the loss in the specific AM within the bar region is progressively more with increasing orbital energy. The physical reason is as follows. The mass loss within the bar region due to the host galaxy is more for a direct orbit when compared with a retrograde orbit with same orbital energy. However, the fraction of mass accumulated within the bar region from the satellite remains similar for a direct and a retrograde orbit (with same orbital energy). This gives rise to the different behaviour in the temporal mass increment for direct and retrograde orbits. 
 
 To conclude, a minor merger event can be a plausible scenario for substantial bar weakening. {\it The efficiency of the bar weakening process during a minor merger event relies on the effectively bringing of the stellar particles from the satellite galaxy within the bar region.} The time-interval of mass accumulation (abrupt versus adiabatic) also plays a pivotal role in disrupting the ordered periodic orbits ($x_1$- and $x_2$- families) which serve as a  backbone of the stellar bar. The vertically extended distribution (as opposed to aligned in disc plane) of accumulated particles from the satellite, with different kinematics as the previous host's disc particles, prevents the sustainability of the bar.

\section{Dependence on Morphology of satellite galaxy}
\label{sec:satellite_dependence}

\begin{figure}
\centering
\includegraphics[width=\linewidth]{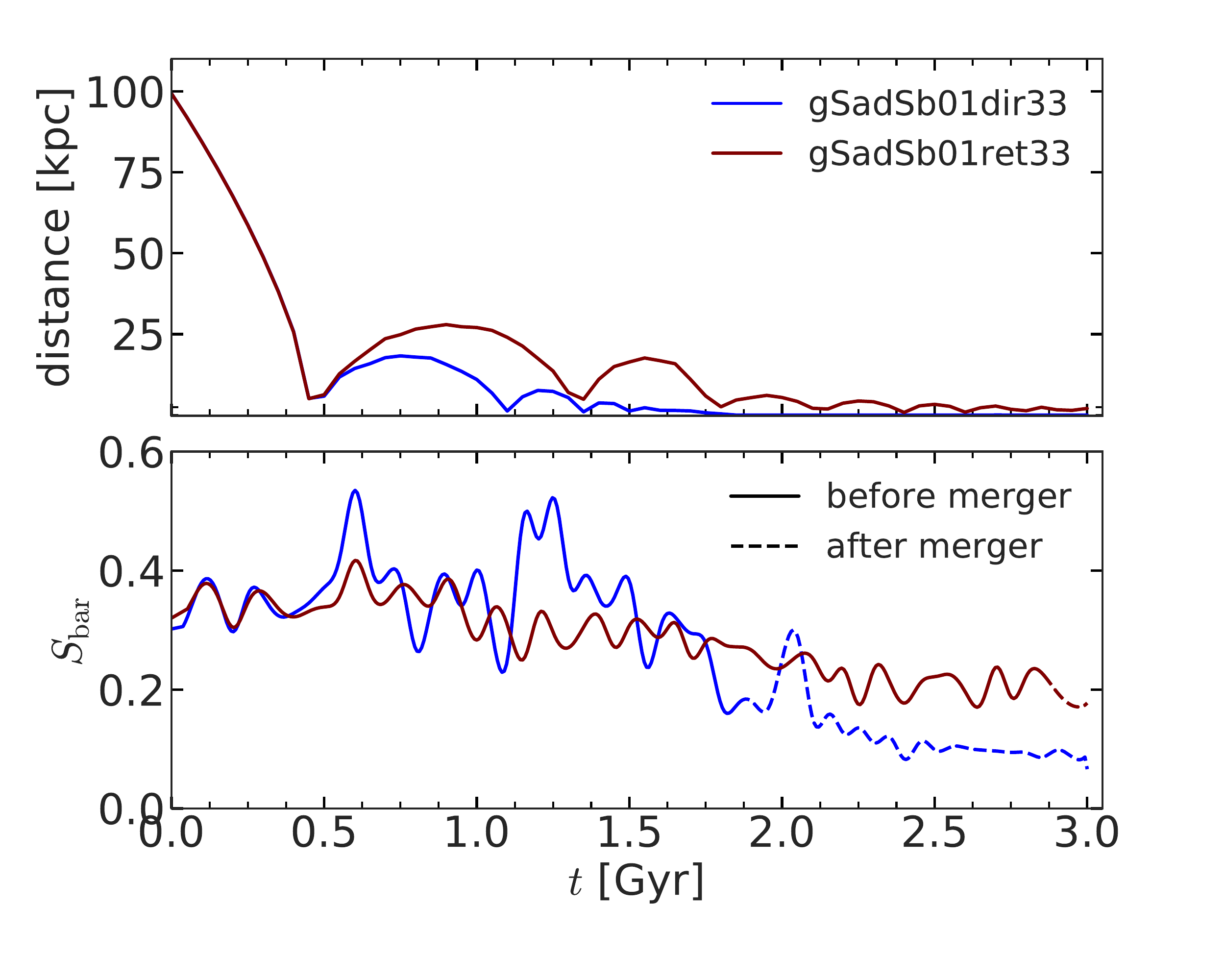}
\medskip
\includegraphics[width=1.05\linewidth]{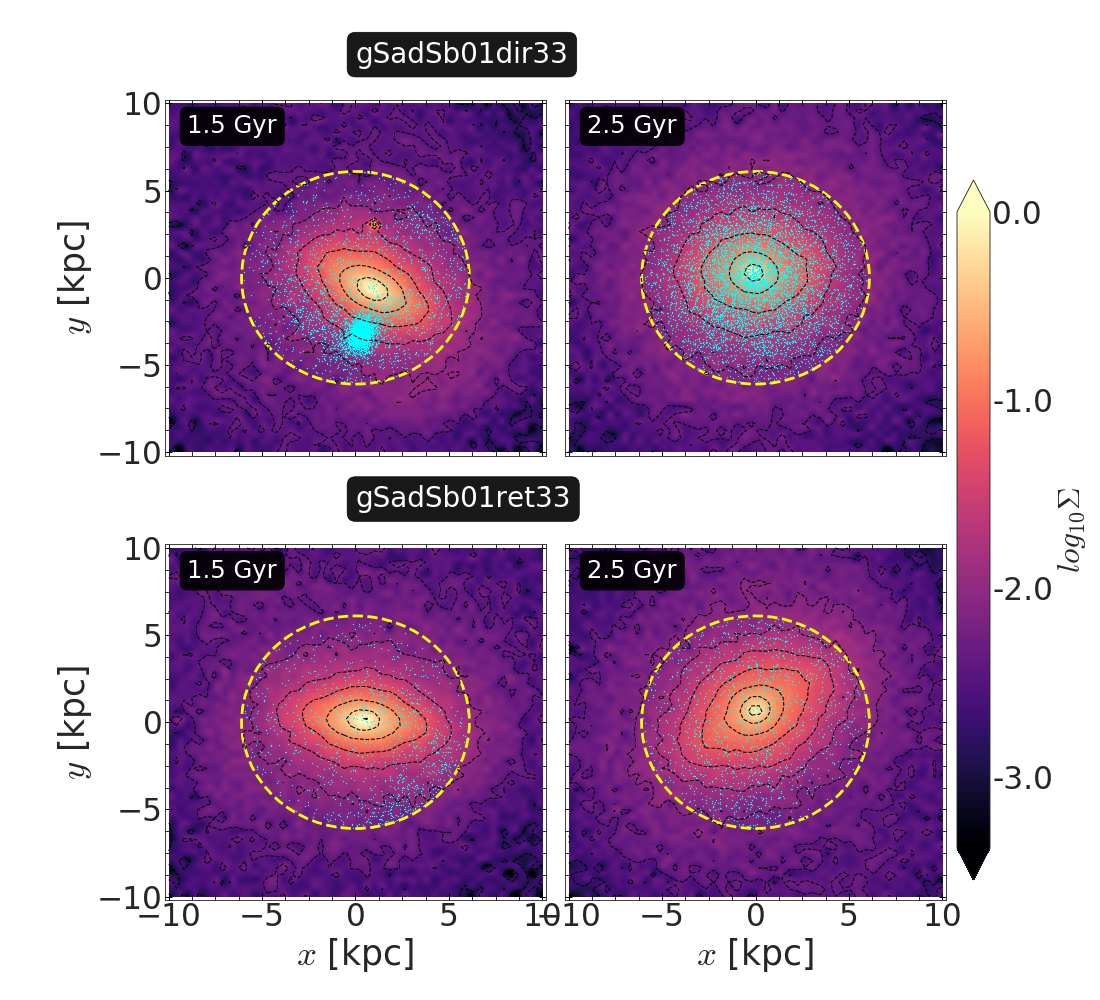}
\caption{{\it Top panel} shows the distance between the centres of the satellite (dSb) and the primary galaxy (gSa) as a function of time for one direct and one retrograde orbit. {\it Middle panel} shows the corresponding temporal evolution of the bar strength ($S_{\rm bar}$) for these two minor merger models. {\it Bottom panels} show the face-on density maps of the host galaxy at different epochs for these two model with satellite's particles (cyan dots) within the initial bar region, over-plotted. The dashed circle (in yellow) indicated the initial bar radius ($R_{\rm bar} = 6.1 \kpc$).}
\label{fig:sat_morphdependence}
\end{figure}

So far, we have considered minor merger scenarios where the satellite galaxy is a dwarf E0 galaxy. Here, we study the efficiency of the bar weakening process when the satellite galaxy has a disc morphology. For this, we consider here two minor merger models where the satellite galaxy is of dwarf Sb-type. Fig.~\ref{fig:sat_morphdependence} ({\it top panels}) show temporal evolution of distance between the centres of mass of these two galaxies and the associated temporal evolution of the bar strength ($S_{\rm bar}$). In both the cases, the bar weakens after the merging happens. However, the degree of bar weakening is different for direct and retrograde orbital configurations. The bar in the direct orbit displays a bar strength $S_{\rm bar} < 0.2$ whereas for the retrograde orbit, the bar strength $S_{\rm bar} \sim 0.2$. 

\begin{figure}
\centering
\includegraphics[width=1.05\linewidth]{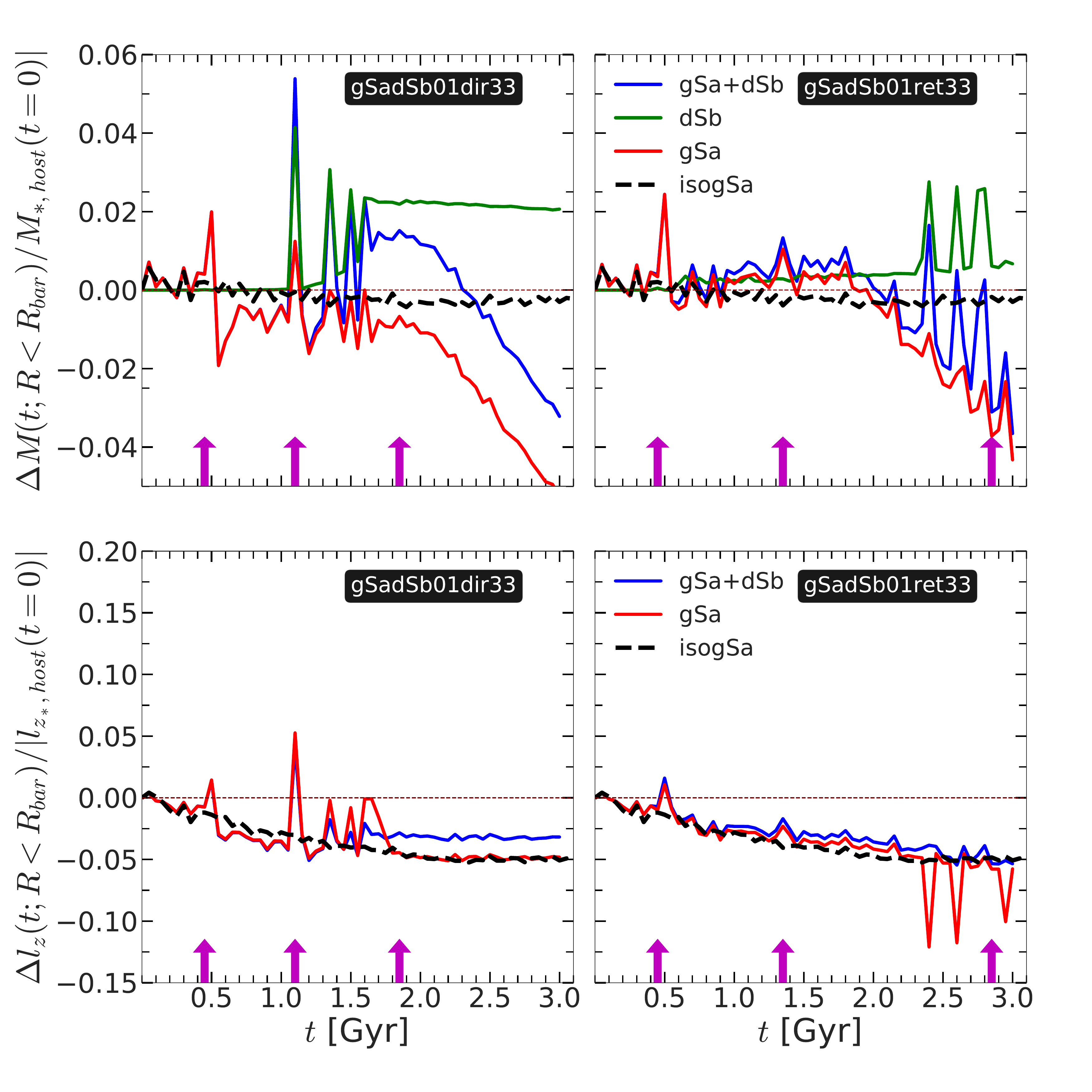}
\caption{{\it Top panels} show the fractional mass change within the bar region while {\it bottom panels} show the corresponding fractional change in the internal specific angular momentum as a function of time, for two minor merger models. The host galaxy is of giant Sa-type and the satellite is of dwarf Sb-type. Black dashed lines denote the same for the isolated evolution of the host galaxy (\textbf{isogSa}). Vertical arrows (in magenta) indicate the epochs of first and second pericentre passages and the epoch of merger, respectively.}
\label{fig:sat_morphdependence_masslz}
\end{figure}

Fig.~\ref{fig:sat_morphdependence} ({\it bottom panel}) also brings out the different scenarios of mass accumulation of stellar particles from the satellite galaxy within the bar region. 
We then quantified the temporal change in the mass and the specific AM content within the bar region for these two models. This is shown in Fig.~\ref{fig:sat_morphdependence_masslz}. The mass accumulation from the satellite within the bar region is less for the retrograde orbit than in direct orbit case. Note that the merger occurs at a later epoch for the retrograde case when compared with the direct orbit. Hence,  by the end of simulation run ($t= 3 \Gyr$), the merger remnant for the retrograde orbit, did not get much time to readjust fully. Also, the addition of satellite's stellar particles (with high specific AM) compensates a part of specific of AM loss due to the host galaxy, within the bar region. This trend is similar to the minor merger models with spheroidal satellite galaxy. However, as the fraction of stellar particles from the satellite galaxy within the bar region is small for the retrograde orbit than the direct orbit, the net change in specific AM for the host plus satellite system closely follow that for the host galaxy.

In the previous sections, the mass accumulation and gain in specific AM in the central bar region are shown to play key roles in the bar weakening. Next, we compare how these processes vary with the morphology of satellite galaxy (discy versus spheroidal). We find that, for the same orbital configuration, the mass accumulation process in the central part from the satellite is more efficient for a spheroidal satellite as compared to a discy satellite galaxy. This happens because for a given orbit, a satellite with higher central concentration is less resistive to the tidal effect of the host galaxy. Consequently it decays rapidly in the central part of the host galaxy.  The variation in accumulated mass fraction with satellite's morphology affects the change in specific AM as well, within the bar region. When the orbital parameters are kept fixed, the gain of specific AM within the central bar region due to the satellite's is lesser for a discy satellite than a spheroidal satellite. Therefore, the net change in specific AM for the host plus satellite system close follow the evolution of specific AM for the host galaxy. This further outlines the importance of effectively bringing the stellar particles from the satellite galaxy within the bar region on the efficiency of the bar weakening process.

\section{Discussion}
\label{sec:discussion}

\subsection{ Case of a delayed merger scenario}
\label{sec:delayedMerger}

\begin{figure*}
\centering
\includegraphics[width=1.05\linewidth]{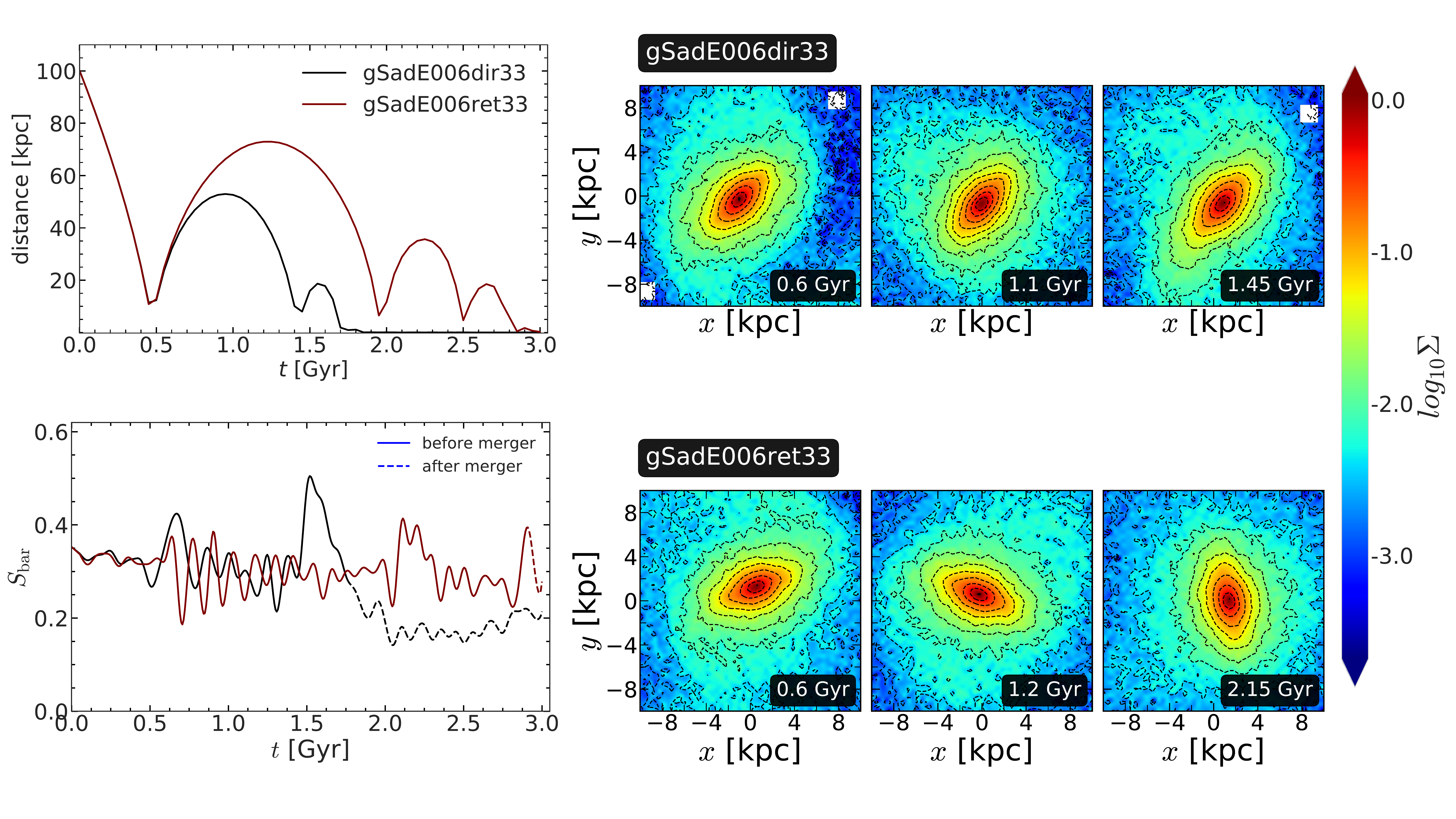}
\caption{{\it Upper left panel} shows the distance between the centres of the satellite (dE0) and the primary galaxy (gSa) as a function of time for two minor merger models (one direct and one retrograde) where merging process is delayed. {\it Bottom left panel} shows the corresponding temporal evolution of the bar strength ($S_{\rm bar}$). {\it Right panels} shows the zoom-in view of face-on density distribution in the central $20 \kpc \times 20 \kpc$ region for the host plus satellite (gSa+dE0) system, at different epochs for these models. Black lines denote the contours of constant surface density. For delayed mergers, bars survive for longer time-scales.}
\label{fig:collage_delayed_merger}
\end{figure*}

Here we consider a scenario of delayed minor merger and study the temporal evolution of the bar properties in the host galaxy. To do that, we considered two minor merger models for which the merger happens at a very later epoch. Thus, these models mimic the scenario of fly-by encounters. Fig.~\ref{fig:collage_delayed_merger} ({\it top left panel}) shows the time evolution of the distance between centres of mass of two galaxies whereas Fig.~\ref{fig:collage_delayed_merger} ({\it bottom left panel} shows the corresponding temporal evolution of the bar strength ($S_{\rm bar}$) for these two models. As seen clearly, after each pericentre passage, the bar strength tend to increase; this increment is more for the direct orbit than the retrograde orbit. The substantial bar weakening happens \textit{only after} the satellite galaxy merges with the host galaxy.  This trend is most prominent for the model {\bf gSadE006ret33} where the merger happens around $t = 2.85 \Gyr$. The face-on density distribution shown at different epochs or the model {\bf gSadE006ret33} clearly display the presence of a prominent stellar bar in the central regions. This is consistent with the past studies, where a stellar bar can be excited due to a fly-by encounter (see references in section~\ref{sec:intro}). This also stresses the crucial role of \textit{mergers} for the bar weakening process as demonstrated in the previous sections. 

\subsection{Bar fraction and gas accretion}
\label{sec:other_issue}

In most of the cases explored here, the satellite or the perturber galaxy merges with the host galaxy; the typical time-scale for the merger to happen is around $2 \Gyr$ after the start of the simulation run. We showed that an initial bar weakens substantially after the merger is completed. Since minor mergers are common in the local Universe \citep[e.g., see][]{Frenketal1988,CarlbergandCouchman1989,LaceyandCole1993,Gaoetal2004,FakhouriandMa2008,Jogeeetal2009,Kavirajetal2009}, therefore the findings of this paper is in apparent tension with the high frequency of the bar incidence in nearby galaxies.  In addition, a galaxy might undergo multiple minor merger events during their entire lifetime  \citep[e.g., see][]{Hopkinsetal2009}; thus, making the bar weakening event more inevitable. However, in reality, the situation is different as a galaxy might be accreting cold gas \citep[e.g., ][]{BirnboimDekel2003,Keresetal2005,DekelBirnboim2006,Ocvriketal2008,Cornuaultetal2018} either during the mergers or at a later stage and this in turn could rejuvenate the bar \citep[e.g., see][]{SemelinandCombes2002,Bournaudetal2005,Combes2008,Marinoetal2011}. Indeed, recent observational studies have pointed out such indication of bar rejuvenation event \citep[e.g.][]{BarwaySaha2020}. Hence, the bar weakening scenario during a minor merger event as shown in this paper is more appropriate for early-type disc galaxies which are in general gas poor \citep[e.g.,][]{YoungScoville1991}. The fraction of galaxies hosting bars decreases from the late-type gas-rich disc galaxies to early-type gas-poor disc galaxies \citep[e.g. see][]{NairandAbraham2010}. The bar fraction in disc galaxies tends to reach their minimum values for the lenticular/S0 galaxies \citep[e.g., see][]{Aguerrietal2009,Butaetal2010,NairandAbraham2010,Barwayetal2011}. Thus, the recent minor merger events which are {\it dry} to a large extent can be a plausible explanation for the absence of strong bars in the early-type galaxies.

\subsection{Other issues}
\label{sec:few_points}
Here, we discuss a few points relevant for this work.

First of all, the bulges present in the GalMer simulations are spherical and initially non-rotating. Previously, \citet{SahaandGerhard2013} investigated the impact of a rotating classical bulge on the bar growth and the kinematics of the resulting boxy/peanut bulge. Their simulation of isolated galaxies showed that the bar strength decreases in cases with rotating bulges but at the end of 3 Gyr (see their Fig.~2), the bar strength with a rotating bulge (one with the maximum rotation) are within 10 per cent of the non-rotating case. In Fig.~\ref{fig:evolu_isolated}, we compared the isolated runs and the one with the minor mergers. It appears that the minor merger events can reduce the bar strength nearly by a factor of two (at the end of 3 Gyr) from what it was in the isolated case. Given our understanding of the impact of rotating bulges on the bar strength, it might not overwhelm the impact of minor mergers. However, a detailed analysis and new simulations need to be run to understand and quantify this effect in the context of minor mergers.

\begin{figure*}
\begin{multicols}{2}
\includegraphics[width=1.05\linewidth]{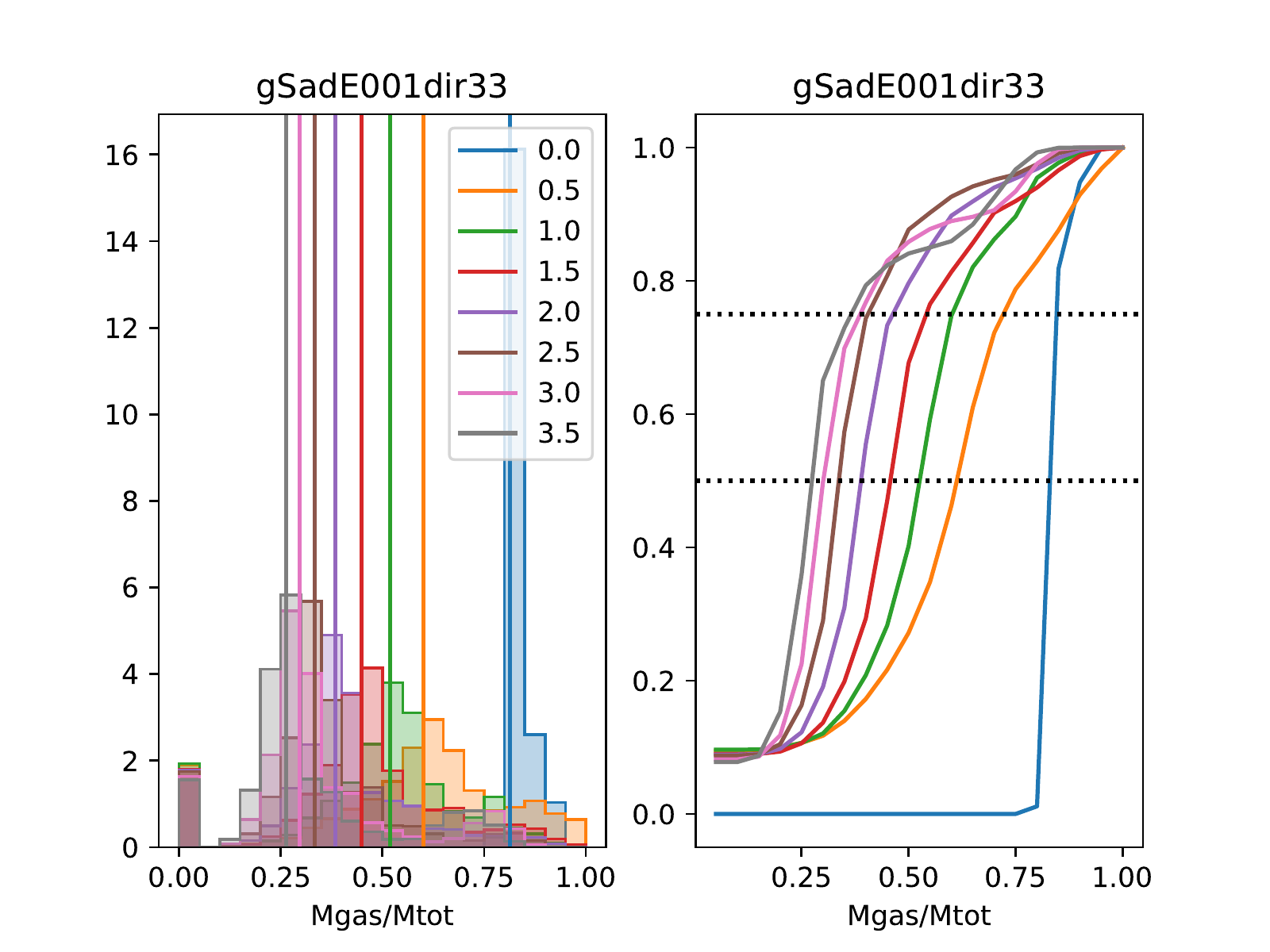}\par
\includegraphics[width=1.05\linewidth]{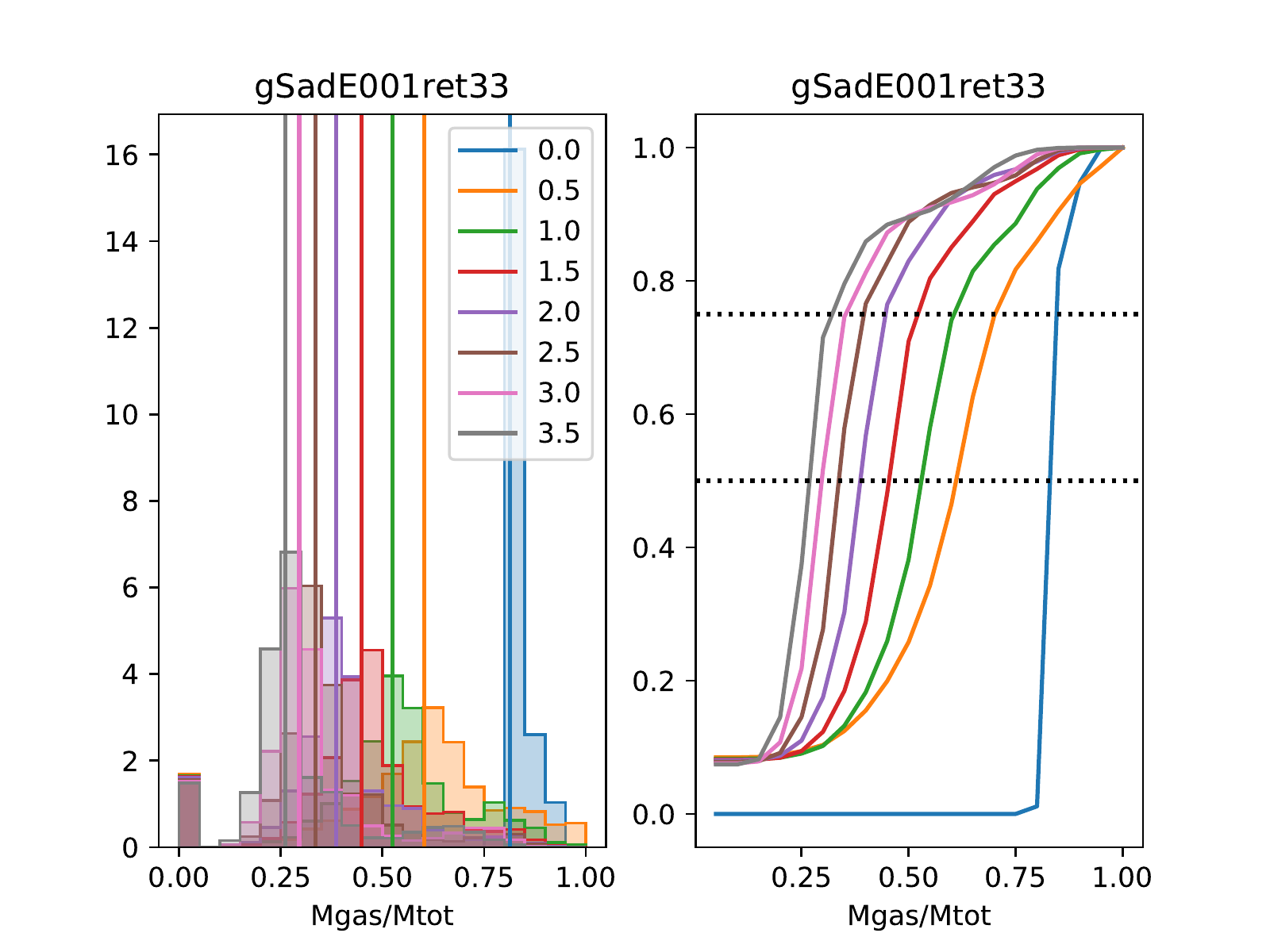}\par
\end{multicols}
\caption{The histogram and the cumulative distribution of gas mass fraction of the hybrid particles within the bar region ($R < R_{\rm bar}$) are shown at different times for minor merger models `gSadE001dir33' (left panels) and `gSadE001ret33' (right panels). The vertical lines correspond to the medians of the distributions at different times. The horizontal black dashed lines in the cumulative distribution correspond to 0.5 and 0.75, respectively. Time (in Gyr) is colour-coded in the legend.}
\label{fig:gas_fraction_tempevolution}
\end{figure*}

Secondly, the GalMer suite of simulations make use of the `hybrid particle' scheme to represent the interstellar gas. This approach of gradually transforming/converting the gaseous to collisionless particles by means of a `hybrid particle' formalism helps to avoid the computational difficulties involved in creating too massive new particles (representative of newly formed stellar population). Furthermore, the numerical models with such a `hybrid particle' scheme were shown to reproduce the Kennicutt-Schmidt law for the interacting galaxies reasonably well \citep[for details see][]{DiMatteoetal2007}.
Fig.~\ref{fig:gas_fraction_tempevolution} shows the temporal evolution of the gas fraction ($M_{\rm gas}/M_{\rm tot}$) of the hybrid particles within the bar region for a direct and retrograde orbital configurations. Here, $M_{\rm tot}$ is the total mass of the hybrid particles and $M_{\rm gas}$ is the mass of its gas component. As seen clearly, the  gas mass fraction within the bar region decreases with time. At later times, around $2 \Gyr$ and after, which are roughly after the merging times of these two minor merger models,  half of the hybrid particles has a gas fraction which is lower than $\sim 30$ per cent within the bar region. In other words, half of the sample is dominated by the stellar component. Moreover, $75$ per cent of the sample is made of particles with a gas fraction less than 0.5 (see the cumulative distributions), which is another indication that for most of the sample, the dynamics is dominated by the stellar component. Hence the impact of the hybrid scheme on the overall dynamics at these late times (which are those where the bar weakening is the strongest) should be minor.

 Lastly, since the softening length is $200 \pc$, so one might wonder whether this resolution can track the evolution of the colder component (gas) in the simulations used here. We mention that, in our models, the hybrid particles represent both the neutral hydroden ($HI$) and the molecular hydrogen ($H_2$). However, the typical spiral galaxies that we simulate have more mass in atomic hydrogen ($HI$) than in $H_2$. While the vertical distribution of $H_2$ is thin, e.g., in the Milky Way, the vertical scale-height of the neutral hydrogen ($HI$) is larger than that of $H_2$, ranging from $500 \pc$ to $1 \kpc$ in the outer parts of the Milky Way \citep[e.g., see][]{Kalberla2009}. Therefore, our models represent it quite well. Thus, it will not have any significant effect on the mains findings of the paper.

\section{Conclusion}
\label{sec:conclusion}
In summary, we investigated the dynamical impact of minor merger of galaxies (mass ratio 1:10) on the survival of a stellar bar, initially present in the host galaxy. We selected a set of minor merger models, with varying orbital energy, orientation of orbital spin vector, morphology of satellite galaxy from the GalMer library of galaxy merger simulation. Then, we studied the temporal evolution of bar properties, before and after the merger occurs. \\
 Our main findings are:

\begin{itemize}

 \item{ A minor merger (mass ratio 1:10) event can substantially weaken the central stellar bar in the merger remnant. The central bar goes through transient bar amplification phases after each pericentre passage of the satellite. The major episode of bar weakening takes place only after the merger happens. This broad scenario holds true for a wide range of orbital parameters considered here.}
 
 \item{Mass accumulation within the bar region from the satellite galaxy plays  a pivotal role in bar weakening process. The freshly added stellar particles from the satellite increases the mass content within the central bar region. The net mass accumulation varies from 3-5 per cent of the total stellar mass of the host galaxy.}
 
 \item{The stellar particles (with high specific AM) from the satellite, accumulated within the bar region after the merger happens, compensates a part of the specific AM loss due to the host galaxy within the bar region. The net loss of specific AM within the bar region for the host plus satellite system is thus less when compared with the specific AM loss solely for the host galaxy.}
 
 \item{The efficiency of accumulation of stellar particles in the central bar region from the satellite depends on the orbital parameters as well as the morphology of the satellite. This, in turn, results in different degree of bar weakening in the minor merger models.}
 
\end{itemize}
The results shown here demonstrates the fact that the minor merger scenario can be a plausible mechanism for the substantial bar weakening.

\section*{Acknowledgement}

We thank the anonymous referee for useful comments which helped to improve this paper. The authors acknowledge support from an Indo-French CEFIPRA project (Project No.: 5804-1). This work makes use of the publicly available GalMer library of galaxy merger simulations which is a part of {\sc HORIZON} project (\href{http://www.projet-horizon.fr/rubrique3.html} {http://www.projet-horizon.fr/rubrique3.html}).

\section*{Data availability}

The simulation data of minor merger models used here is publicly available from the URL  \href {http:/ /galmer.obspm.fr}{http:/ /galmer.obspm.fr}.

\bibliography{my_ref}{}
\bibliographystyle{mnras}


\appendix

\section{Evolution of internal angular momentum of the host galaxy}
\label{appen:int_angmom}

\begin{figure*}
\centering
\includegraphics[width=0.9\linewidth]{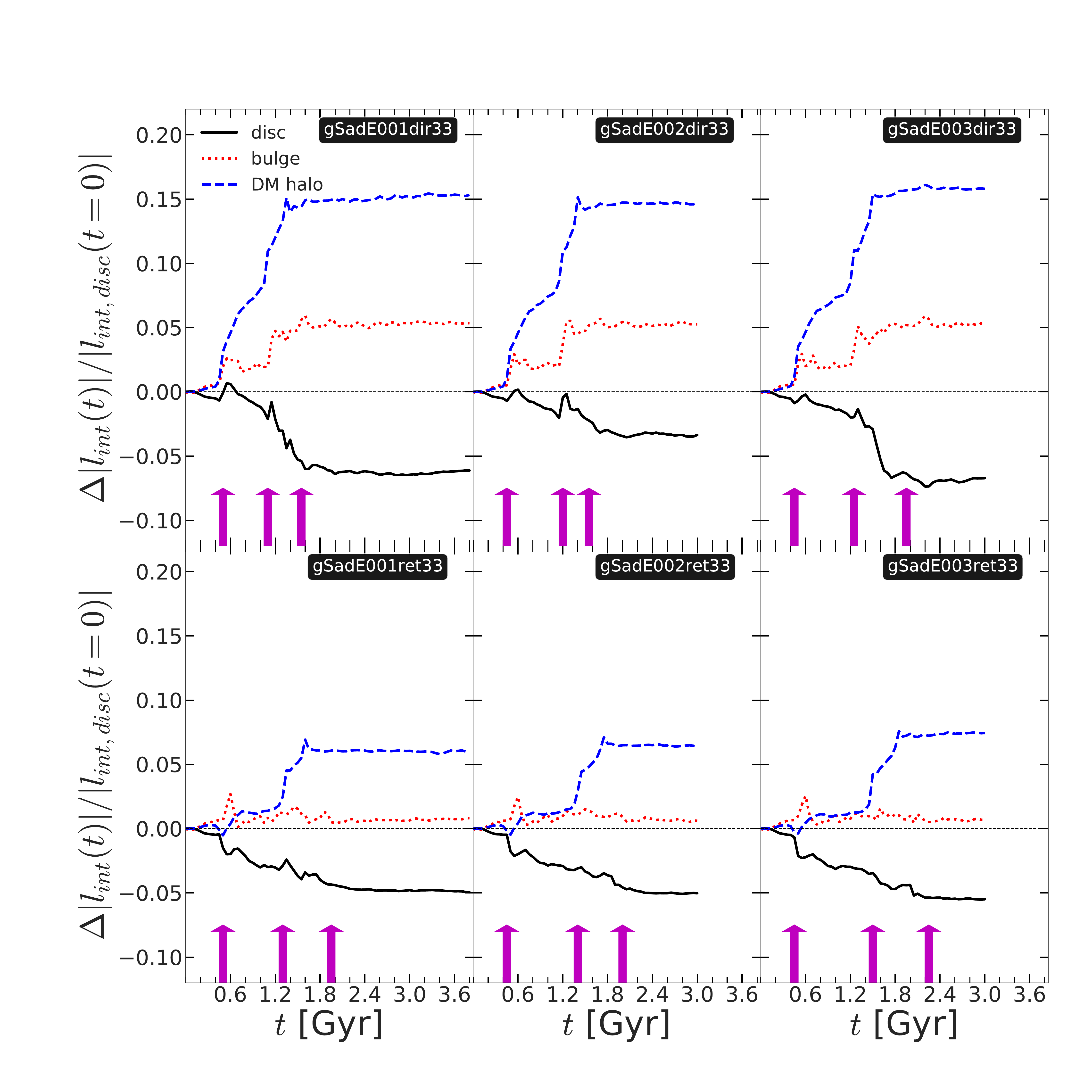}
\caption{Temporal change in internal specific AM shown for different components (disc, bulge, and DM halo) of the host galaxy. {\it Top panels} show for direct orbits whereas {\it bottom panels} show for the retrograde orbits. The averaging is done by the initial disc internal specific AM. Vertical arrows (in magenta) indicate the epochs of first and second pericentre passages and the epoch of merger, respectively. The individual merger models are indicated in each sub-panel.}
\label{fig:internal_specAM}
\end{figure*}

Here, we show how the internal specific AM is getting distributed within different components (disc, bulge, and DM halo) of the host galaxy in a minor merger scenario. At any time $t$, the specific internal AM of the disc component is calculated using $l_{\rm int, d} (t) = \left< \sum_i {\bf r}_{d,i}(t) \times {\bf v}_{d,i} (t)\right> $, where the summation runs over all disc particles of the host galaxy. The specific internal AM for the bulge and the DM halo components are calculated in a similar fashion. The resulting temporal change in specific internal AM for different components is shown in Fig.~\ref{fig:internal_specAM}. 

The disc component loses specific AM, regardless of the orbital energy and orientation of the orbital spin vector. However, the initially non-rotating spherical components, namely, bulge and the DM halo, absorbs part of the orbital AM. While this broad trend holds for all minor merger models shown here, the actual amount of specific AM gain for the bulge and the DM halo components depends on the orbital configuration. To illustrate, the bulge and the DM halo gains more specific AM for a direct orbit when compared with a retrograde orbit with same orbital energy (compare top and bottom panels in Fig.~\ref{fig:internal_specAM}). This trend is in accordance with the findings of \citet{Quetal2010}. The loss of specific internal AM of the disc component is seen to be accompanied by a disc heating phenomenon, causing an increase in the $v/\sigma$ parameter. For the sake of brevity, this is not shown here \citep[for details see][]{Quetal2010}.

\bsp
\label{lastpage}

\end{document}